\documentclass[structabstract]{aa}
\usepackage{graphicx}
\usepackage{float}
\usepackage{txfonts}
\usepackage{aalongtable}
\usepackage{natbib}
\usepackage{multirow}

\bibpunct{(}{)}{;}{a}{}{,}
\begin{document}

\newcommand{\neon}{[Ne~II]}
\newcommand{\lne}{L$_{\rm [Ne~II]}$}
\newcommand{\kms}{km~s$^{-1}$}
\newcommand{\mic}{$\mu$m}
\newcommand{\msun}{M$_\odot$}
\newcommand{\oxy}{[O~I]}
\newcommand{\spit}{{\it Spitzer}}
\newcommand{\lx}{$L_{\rm X}$}
\newcommand{\protop}{protoplanetary disks}
\newcommand{\tts}{T~Tauri~}

   \title{On the origin of \neon~emission in young stars: mid-infrared and optical observations with the Very Large Telescope. \thanks{Based on observations made with ESO Telescopes  Kueyen/UT2 and Melipal/UT3 at the Paranal Observatory under programmes ID 083.C-0471, 084.C-1062, 086.C-0911, and 286.C-5038}}

   \author{Baldovin-Saavedra, C.
          \inst{1,2}
   		  \and Audard, M.
          \inst{1,2}
          \and Carmona, A.
          \inst{1,2}
          \thanks{Now at the CRNS-INSU/UJF-Grenoble 1, Institut de Plan\'eŽtologie et d'Astrophysique de Grenoble (IPAG) UMR 5274, Grenoble, F-38041, France}            
          \and
          G\"udel, M.
          \inst{3}  
          \and 
          Briggs, K.
          \inst{3}            
           \and 
           Rebull, L. M.
           \inst{4}
           \and 
           Skinner, S. L.
           \inst{5}           
           \and 
           Ercolano, B.
           \inst{6,7}
          }

   \institute{ISDC Data Centre for Astrophysics, Universit\'e de Gen\`eve, Chemin d'Ecogia 16, CH-1290 Versoix, Switzerland
         \and
             Observatoire Astronomique de l'Universit\'e de Gen\`eve, Chemin de Maillettes 51, CH-1290 Sauverny, Switzerland
		\and 
			University of Vienna, Department of Astrophysics, T\"urkenschanzstrasse 17, A-1180 Vienna, Austria
		  \and
		   Spitzer Science Center, California Institute of Technology, 220-6 1200 East California Boulevard, Pasadena, CA 91125 USA 
		 \and
		  	Center for Astrophysics and Space Astronomy, University of Colorado, Boulder, CO 80309-0389, USA	
		  \and
		  Universit\"ats-Sternwarte M\"unchen, Scheinerstrasse 1, 81679 M\"unchen, Germany
		  \and 
		  Cluster of Excellence Origin and Structure of the Universe, Boltzmannstrasse 2, 85748 Garching, Germany}

   \date{Received XX; accepted XX}

 
  \abstract
   {The [Ne~II] line 12.81 $\mu$m was proposed to be a good tracer of gas in the environments of proto-planetary disks; its origin is explained by different mechanisms: jets in outflows, photo-evaporative disk winds driven by stellar X-rays/EUV or by the X-ray irradiated proto-planetary disk atmosphere. Previous \textit{Spitzer} studies gave hints toward the neon emitting mechanism by exploring correlations between the line luminosity and properties of the star-disk system. These studies concluded that the origin of the emission is likely related to accretion and outflows, with some influence from X-rays.}
   {We provide direct constraints on the origin of the \neon~emission using high-spatial and spectral resolution observations that allow us to study the kinematics of the emitting gas. In addition we compare the \neon~line with optical forbidden lines.}
   {We obtained high-resolution ground-based observations with VISIR-VLT for 15 stars and UVES-VLT for three of them. The stars were chosen for having bright neon emission lines detected with {\it Spitzer}/IRS. The velocity shifts and profiles are used to disentangle  the different emitting mechanisms producing the \neon~line. A comparison between results from this study and previous high-resolution studies is also presented.}
   {The \neon~line was detected in 7 stars, among them the first confirmed detection of \neon~in a Herbig Be star, V892 Tau. In four cases, the large blueshifted lines indicate an origin in a jet. In two stars, the small shifts and asymmetric profiles indicate an origin in a photo-evaporative wind. CoKu Tau 1, seen close to edge-on, shows a spatially unresolved line centered at the stellar rest velocity, although cross-dispersion centroids move within 10~AU from one side of the star to the other as a function of wavelength. The line profile is symmetric with wings extending up to $\sim \pm 80$ \kms. The origin of the \neon\  line is unclear and could either be due to the bipolar jet or to the disk.  For the stars with VLT-UVES observations, in several cases, the optical forbidden line profiles and shifts are very similar to  the profile of the \neon~line, suggesting that the lines are emitted in the same region. A general trend observed with VISIR is a lower line flux when compared with the fluxes obtained with \spit. We found no correlation between the line full-width at half maximum and the line peak velocity. The \neon~line remains undetected in a large part of the sample, an indication that the emission detected with \spit\  in those stars is likely extended.}
   {}

	
	 \authorrunning{Baldovin-Saavedra et al.}
	 \titlerunning{On the origin of \neon~emission in young stars: VLT observations}

   \maketitle
%
\section{Introduction}
\label{intro_VISIR}

Protoplanetary disks are the outcome of the star formation process. 
They provide the material for the young star and its planetary system. 
The central star indeed accretes matter from the disk until reaching its final mass, after $\lesssim10$~Myr (e.g., \citealt{mannings:1997aa,haisch:2001aa,hillenbrand:2008aa,fedele:2010aa}). 
In order to construct a detailed picture of the process that turns a pre-main-sequence star surrounded by an optically thick disk into a main-sequence star with a planetary system, we need to better understand the disk itself: its composition, size, dynamics, timescales.

The {\it Spitzer} Space Telescope \citep{werner:2004aa} made a large contribution to the studies of protoplanetary disks in the mid-infrared (mid-IR) by revealing the presence of many emission lines from organic molecules \citep{carr:2008aa}, water (\citealt{carr:2008aa,salyk:2008aa,pontoppidan:2010aa}), and atomic species (\citealt{pascucci:2007aa,lahuis:2007aa,flaccomio:2009aa,gudel:2010aa,baldovin-saavedra:2011aa}). 
Among the many emission lines detected by {\it Spitzer}, \neon~12.81 \mic~gained particular interest. It was proposed to be a good diagnostic of gas in the upper layer of the disk, and is used to study the interaction between stellar high-energy irradiation and the disk.
Currently there are three main mechanisms that could explain the presence of \neon~ in the environment of young stars:

\begin{itemize}
\item Irradiated disk atmospheres \citep{glassgold:2007aa}. 
In this scenario, the X-rays from the central star create a warm atmosphere composed of gas in atomic form on top of a cooler molecular layer. \neon~ would be emitted in a region within 20 AU from the central star, the giant planet formation region. 

\item Photoevaporative disk winds driven by EUV (\citealt{clarke:2001aa,alexander:2006aa}), X-rays (\citealt{ercolano:2009aa,owen:2010aa}) or FUV (\citealt{gorti:2009ab}).

\item Shocks (\citealt{hollenbach:1989ab,hollenbach:2009aa}) and jets (\citealt{shang:2010aa}). High velocity outflows from the star interacting with the surrounding material can create strong shocks heating gas to high temperatures. \neon~ would be a good tracer of material at velocities higher than 40-50 \kms.

\end{itemize}

Given the modest spectral resolution of the {\it Spitzer} infrared spectrograph (IRS; R~$\sim 600$, i.e., a velocity resolution of $\sim 500$ \kms; \citealt{houck:2004aa}) the lines detected are unresolved.
Ground-based observations at high spectral resolution are needed in order to determine the origin of the neon-emitting mechanism,
but the number of high-resolution spectra is still small. 
\citet{herczeg:2007aa} detected the [Ne~II] emission line in one out of three spectra of young circumstellar disks observed with the mid-infrared spectrograph on Gemini North (R~$\sim 30000$). 
Based on the measured linewidth  they ruled out an accretion flow origin, favoring the photoevaporative theory. 
\citet{van-boekel:2009aa} presented high resolution ground-based spectroscopy (VISIR-VLT) of the young T Tau triplet, succeeding in spatially separating the  N-S components. These observations showed that the [Ne~II] emission is strongly dominated by outflows heated by shocks. 
\citet{pascucci:2009aa} observed 6 targets with VISIR-VLT. 
For 3 stars in the sample, the line is spectrally resolved, and the line profiles are consistent with those predicted by photoevaporative flow driven by EUV from the central star.
\citet{pascucci:2011aa} performed an extensive study on the neon emission in TW~Hya using VISIR-VLT.
The study confirmed the photoevaporation as emitting mechanism of neon in this star. 
\citet{sacco12} also studied the origin of \neon\ emission with VISIR-VLT. They observed a large sample of young stars and detected the line in 12 out of 32 objects. They concluded that \neon\    emission originates mainly in shocks for Class I protostars. This supports the statistical studies of \citet{gudel:2010aa} and \citet{baldovin-saavedra:2011aa} that jets, if present, tend to dominate the \neon\  emission.
\citet{sacco12} also argued that the emission line stems from the inner disk ($\le 20-40$~AU) for stars with transition and pre-transition disks. Their detailed analysis of the line profiles indicated an origin from a disk wind, although irradiation by EUV/X-rays underestimates the blueshift of the line.
Our study complements the above studies by providing additional VISIR spectra for 15 targets, mainly Class II stars, and further comparing the  \neon\   line with optical lines, in particular forbidden lines such as [O~I], [N~II], and [S~II].

\subsection{Sample}
\label{subs:sample}
\begin{table*}[!ht]
\caption{Stellar properties of the sample studied}
\begin{tabular}{lcccccc}
\hline
Name & $\alpha$~$^{\rm (h~m~s)}$    &  $\delta$~$^{(\circ~\prime~\prime\prime)}$  & Distance $^a$ & Spectral Type & Ref. & L$_{\rm X} ^b $  \\
     &(J2000)                &(J2000) & (pc) &       &  &    (erg~s$^{-1}$)       \\
\hline
 \object{MHO 1}       		& 04:14:26.0   &  +28:06:04.2  &  $140$  & M$2.5$ & 1 & \multirow{2}{*}{$1.6\times 10^{30}$ }  \\
 \object{MHO 2}       		& 04:14:26.4   &  +28:05:59.7  &  $140$  & M$2.5$ & 1 \\
  \object{V892~Tau}    		& 04:18:40.7   &  +28:19:12.3  &  $140$  & B$9$   & 1 & $9.2\times 10^{30}$       \\ 
  \object{CoKu~Tau~1}  		&  04:18:51.6  &  +28:20:26.5  &  $140$  & M$0$   & 1 & ...    \\
  \object{FS~Tau~A}   		&  04:22:02.2  &  +26:57:30.2  &  $140$  & M$0$ &  2 & $3.2\times 10^{30}$       \\
  \object{SST042936+243555} 	&  04:29:36.2  &  +24:35:52.5  &  $140$  & M$3$ & 2 &  ... \\ 
  \object{XZ~Tau}       		&  04:31:39.8  &  +18:13:57.3  &  $140$  & M$2$ & 3 &  $9.6\times 10^{29}$  \\
  \object{L1551~IRS~5} 		&  04:31:34.2  &  +18:08:04.7  &  $140$  & $<$K$6$ & 1 & $2.1\times 10^{28}$  \\  
  \object{UY~Aur}				&  04:51:47.9  &  +30:47:23.0  &  $140$  & M$0$ & 3 & $2.6\times 10^{29}$     \\	
 \object{VZ~Cha}       	    &  11:09:24.4  &  -76:23:22.4  &  $178$  & K$7$ & 4 & $2.5\times 10^{29}$    \\
 \object{RXJ1111.7-7620}		&  11:11:47.0  &  -76:20:10.8  &  $163$  & K$6$   & 5 & $1.5\times 10^{30}$     \\
  \object{V853 Oph}			&  16:28:45.3  & -24:28:18.9   &  $125$  & M$3.75$  & 6 & $3.1\times 10^{30}$       \\ 
  \object{IRS~60}      		&  16:31:31.0  &  -24:24:40.0  &  $125$  & K$2$  & 4  & $2.6\times 10^{29}$      \\
  \object{EC~92}       		&  18:29:57.8  &  +01:12:53.1  &  $260$  & K$7$-M$2$  & 7 & $9.5\times 10^{30}$    \\
\hline
\end{tabular}
\begin{list}{}{}
\item[$^a$]{Distance compiled in \citet{gudel:2010aa}}
\item[$^b$]{L$_{\rm X}$ come from \citet{gudel:2007aa} or \citet{gudel:2010aa}}
\item[References:]{(1) \citet{luhman:2010aa}; (2) \citet{rebull:2010aa}; (3) \citet{hartigan:2003aa}; (4) \citet{torres:2006aa}; (5) \citet{luhman:2004aa}; (6) \citet{wilking:2005aa}; (7) \citet{preibisch:1999aa}}
\end{list}
\label{stellar_prop}
\end{table*}

We recently studied the gas emission with {\it Spitzer}-IRS in a large sample of young stars (\citealt{gudel:2010aa,baldovin-saavedra:2011aa}) obtaining a large number of detections of \neon.
For our dedicated VLT-VISIR high-spectral resolution study, we selected objects accessible by the VLT for which the line fluxes obtained with {\it Spitzer} are higher than $10^{-15}$ erg~cm$^{-2}$~s$^{-1}$, enough for follow-up ground based observations.
The stars in our sample are mainly optically thick disks (Class~II). We included a Class~I object ( \object{L1551 IRS 5}) that has a bright \neon~line detected in the \spit~spectrum, and an intermediate-mass Herbig Be star ( \object{V892 Tau}) because its low resolution spectrum hinted toward neon emission, but just below the detection threshold \citep{baldovin-saavedra:2011aa}.
The stars known to be jet-driving sources are:  \object{CoKu Tau 1} \citep{eisloffel:1998aa},  \object{XZ Tau} \citep{krist:2008aa},  \object{L1551 IRS 5} \citep{rodriguez:2003aa}, and  \object{UY Aur} \citep{hirth:1997aa}.
To our knowledge, the rest of the stars do not have reported outflows or jets in the literature.
 In addition, among the targets selected the following are binaries not resolved by \spit:  \object{MHO-1} and  \object{MHO-2} ($3\farcs9$ separation, \citealt{kraus:2009aa}),  \object{V892 Tau} ($0\farcs05$, \citealt{smith:2005aa,monnier:2008aa}),  \object{FS Tau A} ($0\farcs24$, \citealt{hartigan:2003aa,hioki:2011aa}),  \object{UY~Aur} ($0\farcs88$, \citealt{mccabe:2006aa}),  \object{XZ~Tau} ($0\farcs30$, \citealt{haas:1990aa}),   \object{V853~Oph} ($0\farcs3$, \citealt{mccabe:2006aa}), and  \object{CoKu Tau~1}  ($0\farcs24$, \citealt{padgett:1999aa}).
Table~\ref{stellar_prop} summarizes the stellar properties of the stars selected for this high-resolution spectroscopy follow-up.

The scope of this study is to obtain high-resolution spectra of the \neon~line (12.81355~\mic, \citealt{yamada85}), and by determining the line center and studying the profiles, to obtain observational constraints on the emission mechanism of \neon.
The article is organized as follows: in Sect.~\ref{visir-obs} we describe the VISIR observation strategy and data reduction, in Sect.~\ref{obs-uves} we present complementary observations obtained in the optical with UVES-VLT and its data reduction, in Sect.~\ref{results_visir} we present the results, the discussion in Sect.~\ref{discussion}, and finally the conclusions in Sect.~\ref{conclusions}.

\section{Observations \& Data Reduction}
\label{observations}

\subsection{VISIR-VLT}
\label{visir-obs}

Mid-infrared high-resolution spectra were obtained with the imager and spectrometer VISIR \citep{lagage:2004aa}, installed at the Melipal telescope (UT3) of the VLT in three runs: May 2009, January 2010, and January 2011.
 Spectra were obtained in high-resolution mode (HR) using the long-slit ($32\farcs5$ long). 
This configuration covers a spectral region between $12.793$ and $12.829$~\mic. 
One star (IRS~60) was observed using the cross-dispersed mode (HRX), that uses a short slit of $4\farcs1$, with the same spectral coverage.
The slit width used was $0\farcs4$, achieving a resolving power of $\sim 30 000$ in both configurations,  measured from the FWHM of sky lines present in the spectra, which in velocity resolution corresponds to $\sim 10$~\kms.
The slit was oriented in the default North-South orientation,  except for the sources known to be binaries with a separation such that the single components could be separated with VISIR. 
In those cases the slit orientation was adapted in order to obtain the spectra of the two components of the system.
Standard chop-nodding along the slit was applied to correct for mid-infrared background emission with a chop throw between $8\arcsec$ and $12\arcsec$.
Standard giant stars (from the list of \citealt{cohen99}) or an asteroid (Psyche) were observed immediately before or after the science target to correct for telluric absorption and to obtain the flux calibration.\footnote{In principle, giant stars are not ideal telluric standard stars for medium and high-resolution infrared spectroscopy due to the presence of photospheric absorption lines, and asteroids are better telluric standards. In our case, however, giant stars were sufficient since Ne~II detections (before telluric correction) were very strong, as shown in Fig.~\ref{fig:VISIR_trans}. Photospheric absorption in giant star spectra are faint, after checking with the asteroid spectra. The detections are, therefore, not due to inappropriate telluric correction. Nevertheless, we emphasize that observations to detect  faint Ne~II features should ideally be done with asteroids.}
A summary of the VISIR observations is presented in Table~\ref{sum_obs}, including the date of the observations, the exposure time, the mode (HR high-resolution long-slit or HRX high-resolution cross-dispersed), the airmass at the beginning and end of the observations,  the seeing in the mid-infrared measured from the FWHM of the continuum, the position angle (following the standard convention of positive angles in the North-East direction), the star or asteroid used as calibrator, its respective exposure time,  and airmass.

\begin{table*}[!ht]
\caption{Summary of the VISIR observations}
\begin{tabular}{lcclcccccc }
\hline
Name &  Date  & t$_{\rm on-source}$ &Mode & Airmass & MIR seeing &Pos. Angle & Calibrator (Sp. T) & t$_{\rm cal}$ & Airmass\\
       &      &      (s)      &               &   start/end   &  $(\arcsec)$ & $(\degr)$ &    & (s)    &    \\
\hline
  \object{MHO 1/2}       		&   2011-01-21  & 1800  & HR &  1.71/1.89  & 0.39 &$-153.8$  &HD~23319 (K2.5III) & 240 & 1.02\\
  					&  2011-01-23  & 1800  & HR 	  &  1.68/1.81  &  0.33 &$-153.8$  &HD~29291 (G8III) & 240 & 1.01\\
 \object{V892~Tau}   		&   2010-01-06 &  3600  & HR   &  1.71/1.70  & 0.36 &0  & HD~20644 (K4III) & 240 & 1.69\\ 
  \object{CoKu~Tau~1}  		&   2010-01-04 &  3600  & HR  &  1.72/1.69  & 0.32 &0  & HD~17361 (K1.5III) & 240 & 1.70\\
  \object{FS~Tau~A}    		&   2010-01-04  &  2760  & HR &  1.73/2.14   & 0.33 &0  & HD~27639 (M0III) & 240 & 1.95\\
		    			&   2010-01-05  &  3600  & HR &  1.71/1.62  & 0.29 &0  & HD~27482 (K5III) & 240 & 1.70\\
  \object{SST042936+243555} 	&   2011-01-21  &  1800  & HR &  1.78/2.17  & 0.39 & 0  & HD~27639 (M0III) & 600 & 2.46\\
  \object{XZ~Tau}       		&   2011-01-22  &  1800  & HR &  1.52/1.78  &  0.32 &$-138.6$  &Psyche & 120 & 1.85\\
  \object{L1551~IRS~5}		&   2010-01-05  &  3600  & HR &  1.41/1.78  &  0.28 &0 & HD~27482 (K5III) & 240 & 1.64\\   
  \object{UY~Aur}				&   2011-01-22  &  1800  & HR &  1.76/1.81  &  0.36 &132.4  & HD~29291 (G8III) & 240 & 1.00 \\	
					    &   2011-01-23  &  1800  & HR &  1.88/2.14  &  0.38 &132.4  & Psyche & 120 & 1.77 \\	
  \object{VZ~Cha}       	    &   2011-01-22  &  1800  & HR &  1.87/1.76  & 0.32 &0 & Psyche & 120 & 1.85 \\
  \object{RXJ1111.7-7620}		&   2011-01-21  &  960  & HR &  1.79/1.74  &  0.39 &0  & HD~27639 (M0III) & 600 & 2.46\\
  \object{V853 Oph}			&   2009-06-01  &  6120  & HR &  1.01/1.50  & 0.27 &0  &HIP~90185 (B9.5III) & 600 & 1.11\\ 
  \object{IRS~60}      		&   2009-05-31  &  3600  & HRX  &  1.19/1.01  & 0.40 &0  & HIP~84012 (A2IV) & 840 & 1.05  \\
		    			&   2009-06-01  &  1800  & HRX  &  1.06/1.00  & 0.40 &0  & HIP~84012 (A2IV) & 480 & 1.04 \\
  \object{EC~92}      		&   2009-05-31  &  4080  & HR &  1.15/1.18  &  0.26 &$-173.7$  & HIP~99473 (B9.5III) & 1200 & 1.15\\
		    			&   2009-06-01  &  3600  & HR &  1.36/2.24  & 0.32 &$-173.7$  & HIP~84012 (A2IV) & 840 & 1.38 \\
\hline
\end{tabular}
\label{sum_obs}
\end{table*}

The VISIR raw frames were processed with the standard VISIR data reduction pipeline version 3.7.2, using the command line application {\it esorex}.
The output of the pipeline is a series of files including the extracted spectrum, a 2D image spectrum, a pixel-to-wavelength map, and a synthetic model spectrum of the calibration star.
Starting from the 2D spectrum we have performed further steps based on \citet{boersma:2009aa} for the spectra obtained in HR mode.

{\it i)} Any remaining background emission is corrected by taking each pixel row and fitting a second order polynomial, ignoring the source power spread function (PSF) profile. The result from the fit is then subtracted from each pixel row, obtaining a 2D spectrum that is background-corrected.

{\it ii)} A weight map is created by collapsing the image in the dispersion direction, normalizing it, and then expanding it in the spatial direction. The science frame is then multiplied by the weight map. The spectrum is obtained by collapsing the resulting image in the spatial direction.

{\it iii)} The wavelength calibration is performed using the atmospheric emission lines. 
This is done by taking a raw frame and cross-correlating it with a model of the atmospheric spectrum \footnote{Available on the VISIR Tools website, see 
\texttt{http://www.eso.org/sci/facilities/paranal/instruments}}.
Gaussian profiles are fitted to the atmospheric lines in order to obtain accurately the position of their center.
By fitting a second order polynomial the pixel-to-wavelength conversion is obtained.
This method gives a precision of $0.4-0.6$~\kms~in the wavelength solution.

{\it iv)} The observed calibrator spectrum is corrected for differences in airmass and air pressure with the science target, following the procedure described in \citet{carmona:2011aa}.
The calibrator synthetic spectrum is divided by the observed calibrator spectrum. 
This result is then multiplied to the extracted spectrum of the science target to correct for telluric absorption and to get an absolute flux calibration.
The uncertainty in the flux calibration is of the order of $20-30\%$. 

Steps {\it i}, {\it ii}, and {\it iii} were applied to both the science target and its calibrator. The VISIR spectra were corrected for barycentric and radial velocity.

Whenever an asteroid was observed to correct for telluric absorption, the absolute flux calibration was based in \spit~fluxes.
In the particular case of UY~Aur, the \spit~flux corresponds to the binary, therefore to obtain an absolute flux calibration for the single components we used in addition the flux ratio between the components, derived from high-resolution imaging in the N band \citep{mccabe:2006aa}. 

For the binaries, we converted the separation of the spectra from pixels to distance in the sky using the VISIR pixel scale ($0\farcs127/$pix).
The binary systems resolved by our VISIR observations are  \object{MHO-1/2}, and  \object{UY~Aur}, for which the projected separations observed are consistent with the separations reported in the literature.
The small separation of  \object{XZ~Tau} did not allow the system to be spatially resolved, although the observed cross-profile was broad. The  \object{CoKu Tau 1} and  \object{FS Tau A} binaries are also not resolved and show  cross dispersion profiles consistent with a single source.
Observations of  \object{EC~92} include the nearby star EC~95, located at a distance of $5\arcsec$, however the spectrum of EC~95 has a very low signal-to-noise ratio and it is not exploitable.

\subsection{UVES-VLT}
\label{obs-uves}

 After the end of our last VISIR run we requested observing time to obtain optical spectra with UVES-VLT for a number of targets with detected \neon~emission in VISIR, under DDT program 286.C-5038(A).
The purpose was to derive radial velocities at high precision, only possible from high-resolution spectra (see details in \ref{app:rad_vel_cal}).
Radial velocities are obtained by measuring the shift of the absorption lines of the stellar atmosphere with respect to a reference spectrum and are crucial to determine the rest velocities of the \neon~lines detected with VISIR.
UVES (Ultraviolet and Visual Echelle Spectrograph) is located at the Nasmyth platform B of the second Unit Telescope (Kueyen) of the VLT, offers cross-dispersed echelle spectra between 300 up to 1100 nm. 
The light beam is splitted into two arms; UV-Blue and Visual-Red. These two arms can operate separately or in parallel.
We used the red arm covering the wavelength range $500-595$ and $605-700$~nm. 
This setting allowed us to cover a region rich in absorption lines, required for the determination of the stellar radial velocity,
and including important emission lines, such as \oxy~and H$\alpha$.
The slit width used was $1\farcs2$, achieving a resolving power $R\sim 33 000$.
The summary of the observations is presented in Table~\ref{sum_obs_uves}.

\begin{table}
\begin{center}
\caption{Summary of the UVES observations}
\begin{tabular}{lcccccc}
\hline
 Name    & Date  & Exp. time & Airmass & Seeing \\
       &     &      (s) & & $(\arcsec)$  \\
\hline
  \object{CoKu~Tau~1} & 2011-03-02  & 900 & 2.0 & 1.4/1.2\\
  \object{FS~Tau~A}   & 2011-03-02  & 900 & 2.1 & 1.2/1.3\\
  \object{V853 Oph}   & 2011-03-02  & 300 & 1.3 & 1.3/1.1\\ 
\hline
\end{tabular}
\label{sum_obs_uves}
\end{center}
\end{table}

We reduced the data following the standard UVES pipeline recipes, version 4.9.0, using the command line application {\it esorex}.
Optical data need to be corrected for several instrumental effects.
To correct for the electronic noise of the camera and possible systematics, a short exposure (bias) is taken with the shutter closed; the bias frames give the read out of the CCD detector for zero integration time. 
Several bias frames are combined into a master bias; in the case of UVES five bias frames are taken. The more bias exposures are taken, the less noise will be introduced into the corrected images.
Furthermore, as the telescope does not illuminate the detector homogeneously, and the quantum efficiency of the CCD is not necessarily the same over all the pixels, an exposure of an homogeneously illuminated area (flat field) needs to be taken. Five flat fields are then combined to create the normalized master flat field for UVES observations.
The bias frame is subtracted of each raw science frame and the resultant spectrum is divided by the master flat field frame.
In general, a correction for the dark current of the CCD is also needed, and this is done by taking a long exposure with the shutter closed (dark frame). 
In the case of UVES, the detector dark currents can be considered negligible and be excluded from the data reduction process. 
The wavelength calibration of UVES spectra is done by using observations of a ThAr lamp and is part of the standard procedures of the UVES pipeline.
 The error in the determination of the wavelength solution is $2$~m$\AA$, typical of the UVES pipeline. The spectra were also corrected for radial and barycentric velocity.
Given that we did not observe standard stars, the correction for the telluric absorption features and absolute flux calibration was not possible.
However, the data reduction process includes the correction of the spectra for the sky airglow emission. We used the $R$ magnitude and Johnson zero point photometry to determine
the flux density in the continuum.

\section{Results}
\label{results_visir}

\subsection{\neon~detections in VISIR spectra}

\begin{figure*}
\begin{center}
\includegraphics[width=.99\textwidth]{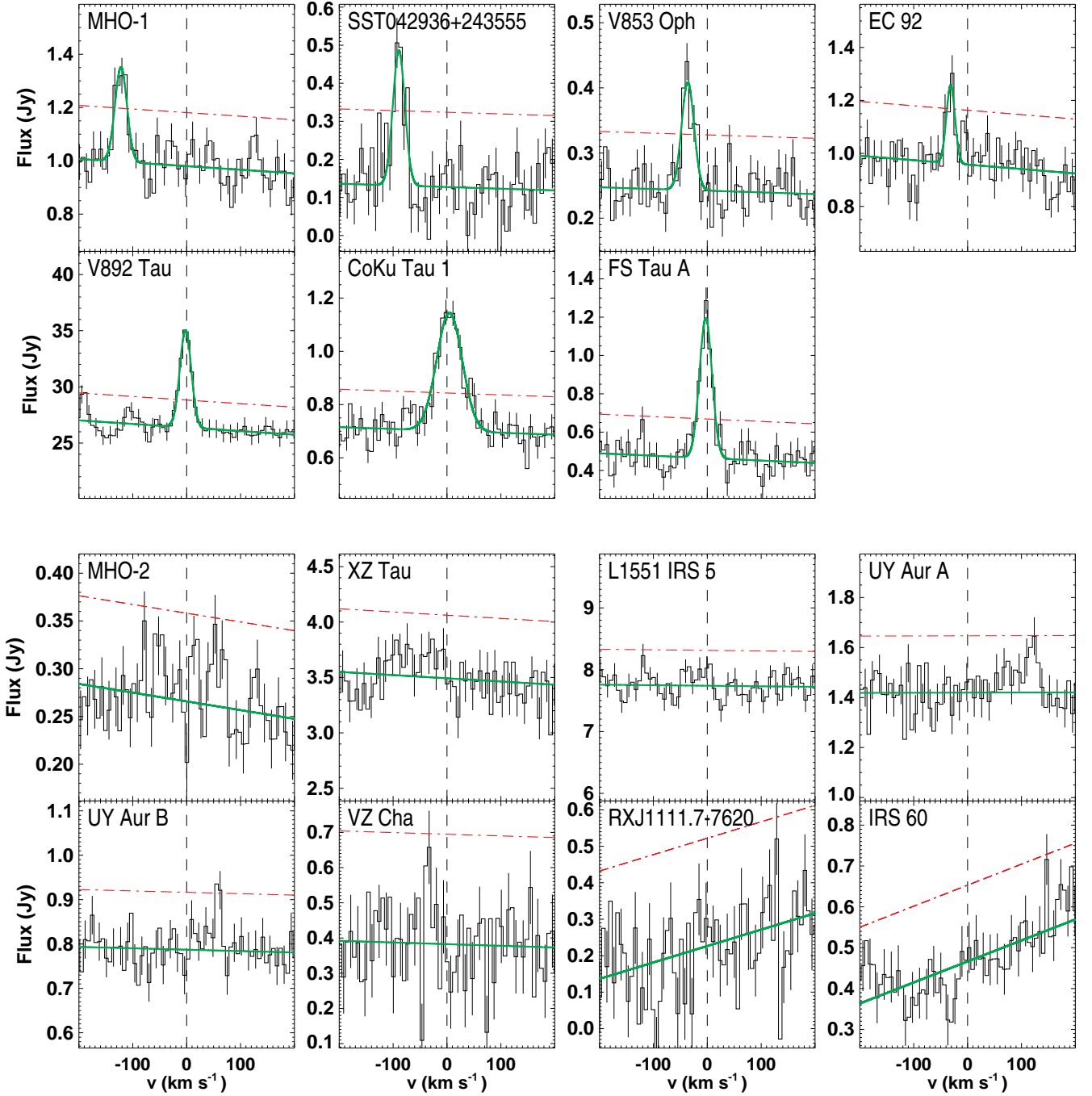}
\end{center}
\caption[Spectra of the stars observed with VISIR]{Spectra of the stars observed with VISIR plotted in stellocentric frame and rebinned by a factor of two (average spectra are shown when there are more than one observations). The detections are plotted in the upper panel, non detections in the lower one. The green line is the Gaussian fit to the line, the red dot-dashed line is the $3\sigma$ detection threshold per pixel. The error bars shown are $1\sigma$ values per pixel.}
\label{fig:visir_ne2_spect}
\end{figure*}

The \neon~emission line was detected in seven stars in our sample:  \object{MHO-1},  \object{V892~Tau},  \object{CoKu~Tau~1},  \object{FS~Tau~A},  \object{SST~042936+243555},  \object{V853~Oph}, and  \object{EC~92}.
The line was fitted using a Gaussian profile plus a linear component to account for the continuum.
All the parameters of the function were left free to vary.
For the cases with non detections, an estimation of the upper limit was calculated as three times the standard deviation of the continuum flux ($\sigma$) times the spectrograph full width half maximum (FWHM).
Figure~\ref{fig:visir_ne2_spect} displays the spectra after wavelength and flux calibration, plotted including the error bars in the flux. 
The green line represents the Gaussian fit, and the red dot-dashed line is the $3\sigma$ upper limit threshold.
The spectra are plotted in the stellocentric frame; the vertical dashed line shows the position the lines would have if centered at the stellar radial velocity. The figure was divided in two panels: detections in the upper panel and non detections in the lower one. In addition, the detections are separated in two groups: stars showing a line blueshifted to high velocities, and stars with a line nearly centered or shifted by a few km~s$^{-1}$.   
 Figure~\ref{fig:VISIR_trans} displays the observed spectra of the stars and its telluric standard before applying any velocity correction.
Only one data set is included for the stars observed in two different days.

Table~\ref{table:fit_results} shows the results obtained from the fits to the \neon~emission line: the integrated line fluxes or upper limits in Col. $2$, the FWHM of the detected lines in Col. $4$, and the center of the Gaussian in Col. $5$.
In addition, we report the {\it Spitzer} fluxes for comparison in Col. $3$, the stellar radial velocity and its reference in Cols. $6$ and $7$, respectively. The errors of the measurements are reported between parentheses.

\begin{table*}
\caption{Results for the \neon~line obtained from the VISIR spectra. The peak is given in the stellocentric frame. Errors are reported between parentheses.}
\begin{tabular}{lcccccl}
\hline
 $[1]$ & $[2]$& $[3]$ & $[4]$ & $[5]$ & $[6]$ & $[7]$ \\ 
Name &  Flux &  {\it Spitzer} Flux $^\clubsuit$& {\rm FWHM} & Peak$^\dagger$ & v{\tiny rad} & v{\tiny rad} Ref.\\
   & ($10^{-15}$ erg~cm$^{-2}$~s$^{-1}$) & ($10^{-15}$ erg~cm$^{-2}$~s$^{-1}$)  & (km~s$^{-1}$)  &  (km~s$^{-1}$) & (km~s$^{-1}$) & \\ 
\hline
\multicolumn{7}{c}{Detected lines} \\
\hline
MHO 1            & $8.0~(1.5)$   & $120~^{\diamondsuit}$   & $26.7~(3.3)$  &  $-121.8~(6.6)$	& $16.0~(6.4)^{\spadesuit}$ & \citet{bertout:2006aa}\\
V892~Tau         & $179~(16)$    & $<180$  & $26.0~(3.4)$  &  $-2.13~(6.5)$       & $16.0~(6.4) ^{\spadesuit}$ &{\citet{bertout:2006aa}}\\ 
CoKu~Tau~1      & $20.4~(2.5)$  & $120$   & $55.2~(3.3)$  &  ${3.6~(1.5)}$       & $15.0~(0.8)$ & {\citet{white:2004aa}}\\
FS~Tau~A        & $16.6~(1.7)$  & $77$    & $26.8~(1.7)$  &  $-2.9~(0.7)$		& $17.1~(0.1)$ & This work \\
SST042936+243555& $7.8~(1.7)$   & $11$    & $26.0~(3.8)$  &  $ -89.1~(6.6)$    & $16.0~(6.4) ^{\spadesuit}$ & \citet{bertout:2006aa}\\
V853 Oph	     & $3.7~(0.8)$   & $16$    & $26.5~(3.9)$  &  $-35.8~(1.6)$     & $-8.9~(0.1)$ & This work\\ 
EC~92           & $4.2~(1.2)$   & $15$    & $16.2~(4.3)$  &  $-31.3~(2.3)      $& $-7.1~(1.5)$	&  \citet{covey:2006aa} \\

\hline
\multicolumn{7}{c}{Upper limits} \\
\hline
MHO 2   			&  $<~0.5$ & $120~ ^{\diamondsuit}$  & ...   & ...   & $16.0~(6.4) ^\spadesuit$& \citet{bertout:2006aa}\\
XZ~Tau       	& $<~5.2$  & $32$   & ...   & ...   & $17~(7)$ & \citet{folha:2000aa}\\
L1551~IRS~5 	    &  $<~4.3$ & $580$  & ...   & ...   & $8.2~(1.5)$ & \citet{covey:2006aa} \\   
UY~Aur~A		&  $<~2.4$ & $49~ ^{\diamondsuit}$   & ...   & ...   & $18~(3)$ & \citet{barbier-brossat:1999aa}\\	
 UY~Aur~B	    &  $<~1.2$ & $49~ ^{\diamondsuit}$  & ...   & ...   & $18~(3)$  & \citet{barbier-brossat:1999aa}\\	
 VZ~Cha          & $<~2.3$ & $3.8$ & ... & ... & $19.0~(1.7)$ & \citet{torres:2006aa} \\
 RXJ1111.7-7620 &  $<~3.3$  & $5.1$ & ... & ... & $15.5~(0.3)$ & \citet{torres:2006aa}\\
 IRS~60         &  $<1.6$ & $14$ & ... & ... &  $-2.2$   & \citet{torres:2006aa} \\
\hline
\end{tabular}
\begin{list}{}{}
\item[$\clubsuit$]{{\it Spitzer} fluxes come from \citet{gudel:2010aa} or \citet{baldovin-saavedra:2011aa}}
\item[$\dagger$]{The error in the center was calculated considering the contribution of the error in the Gaussian fit and in the stellar radial velocity}
\item[$\spadesuit$]{In these cases there is no radial velocity in the literature, the average radial velocity of Taurus was assumed.}
\item[$\diamondsuit$]{The binary systems UY~Aur (A,B) and MHO~1/2 cannot be separated in the {\it Spitzer} spectrum, therefore the flux reported is for the system.}
\end{list}
\label{table:fit_results}
\end{table*}

\begin{figure*}[ht]
\begin{center}
\includegraphics[width=\textwidth]{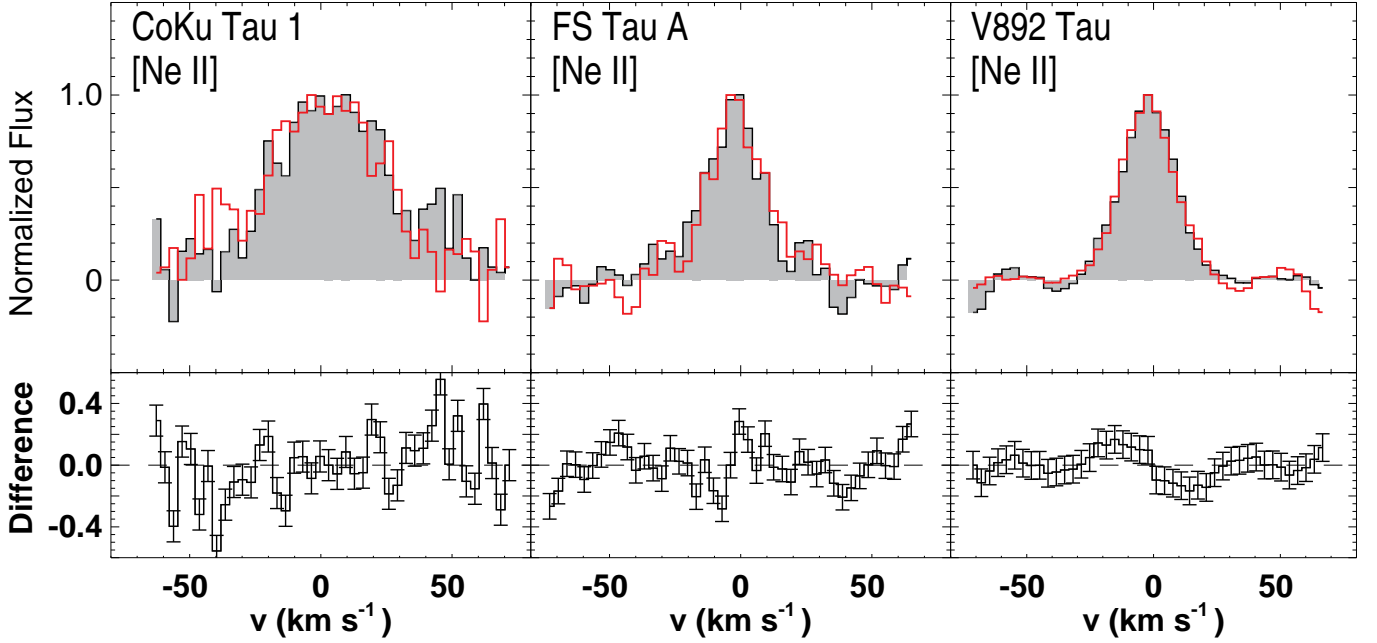} 
\end{center}
\caption{Upper panel: Plotted as grey filled area the profile of the \neon~line and in red solid line the profile flipped. Lower panel: Difference between the original and the flipped profile.}
\label{fig:ne2_profiles}
\end{figure*}

The detection of [Ne II] from the Herbig Be star  \object{V892 Tau} that we present here is the first report of [Ne II] emission from an intermediate mass pre-main sequence star.
The \neon~line is detected close to the stellar rest velocity in three cases:  \object{CoKu~Tau~1},  \object{FS Tau A}, and  \object{V892 Tau}. 
For them, we studied the symmetry of the line profiles by using the procedure described in \citet{pascucci:2011aa}; each line is flipped and shifted in wavelength direction to get a good match with the original profile. 
The two profiles are then subtracted and the result is plotted; the more symmetric the line, the closer to zero the difference.
The results are shown in Figure~\ref{fig:ne2_profiles}: in the upper panel of the figure, the original and the flipped spectrum are shown, while the lower panel displays the difference between both profiles.
{This method does not show any evident asymmetry in the line profiles.}

\begin{figure*}[!ht]
		\begin{minipage}[!t]{\textwidth}	
	\centering
     		$\begin{array}{ccc}
	    			\includegraphics[width=5.6cm]{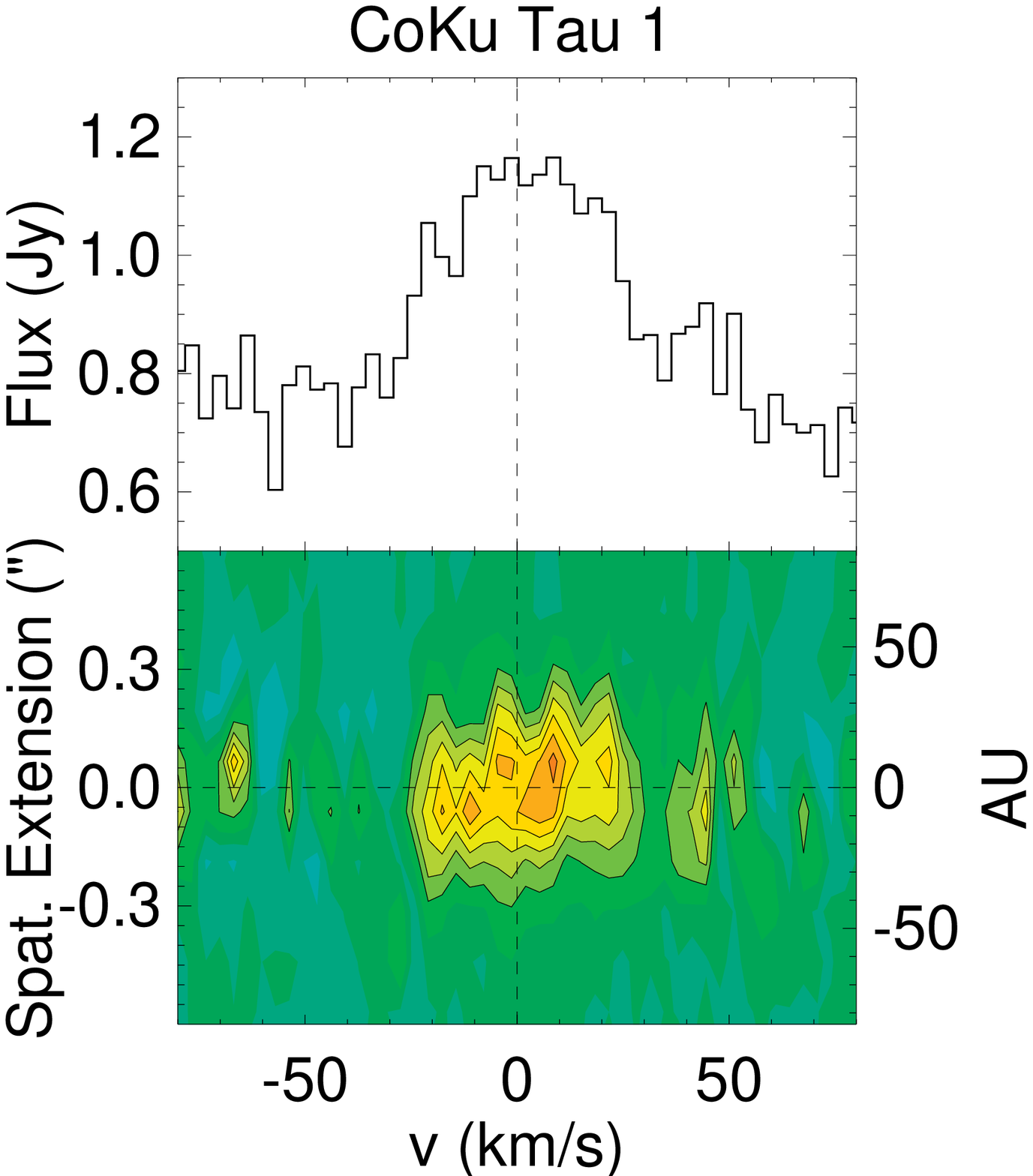} &
	    			\includegraphics[width=5.6cm]{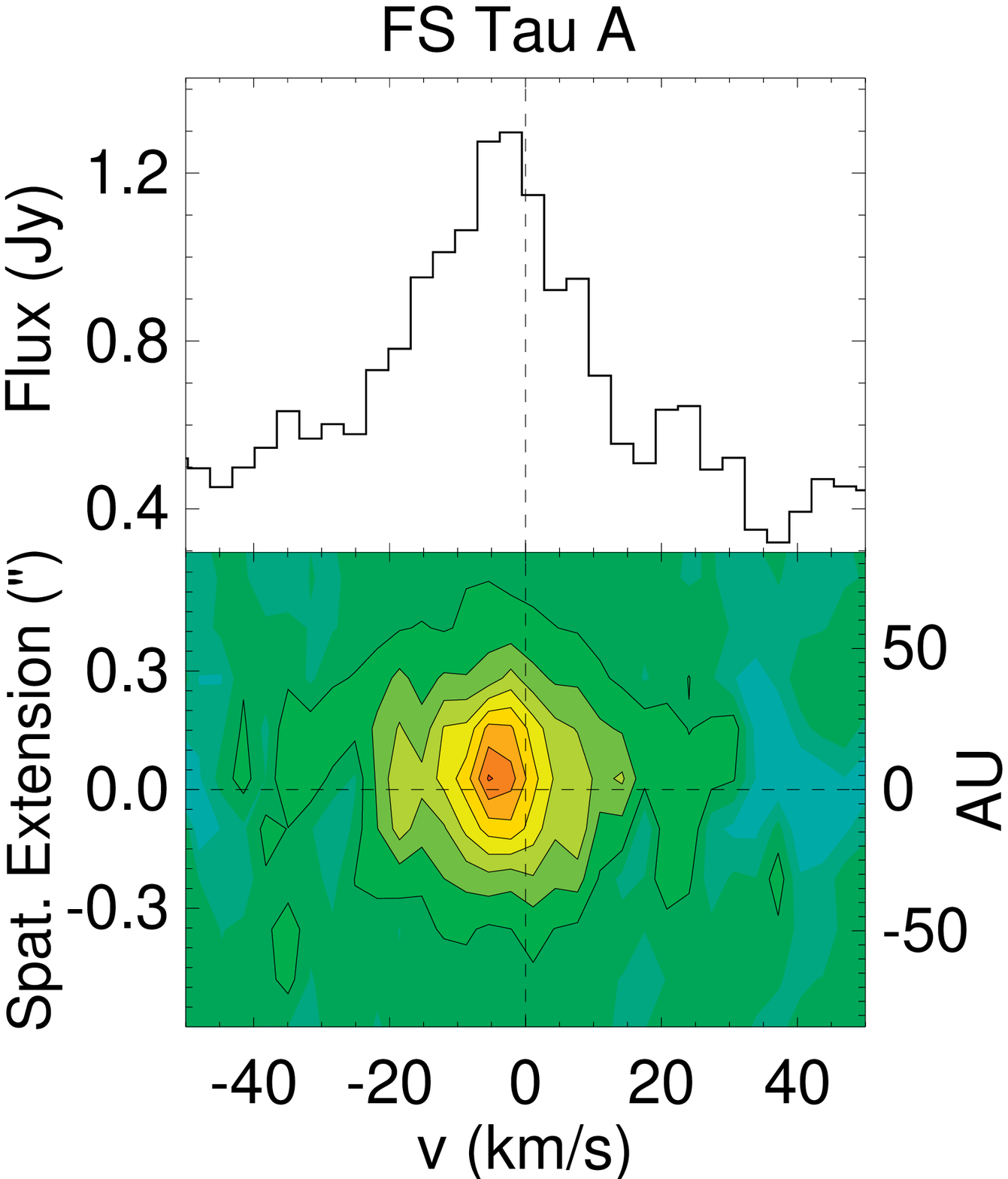} &
				\includegraphics[width=5.6cm]{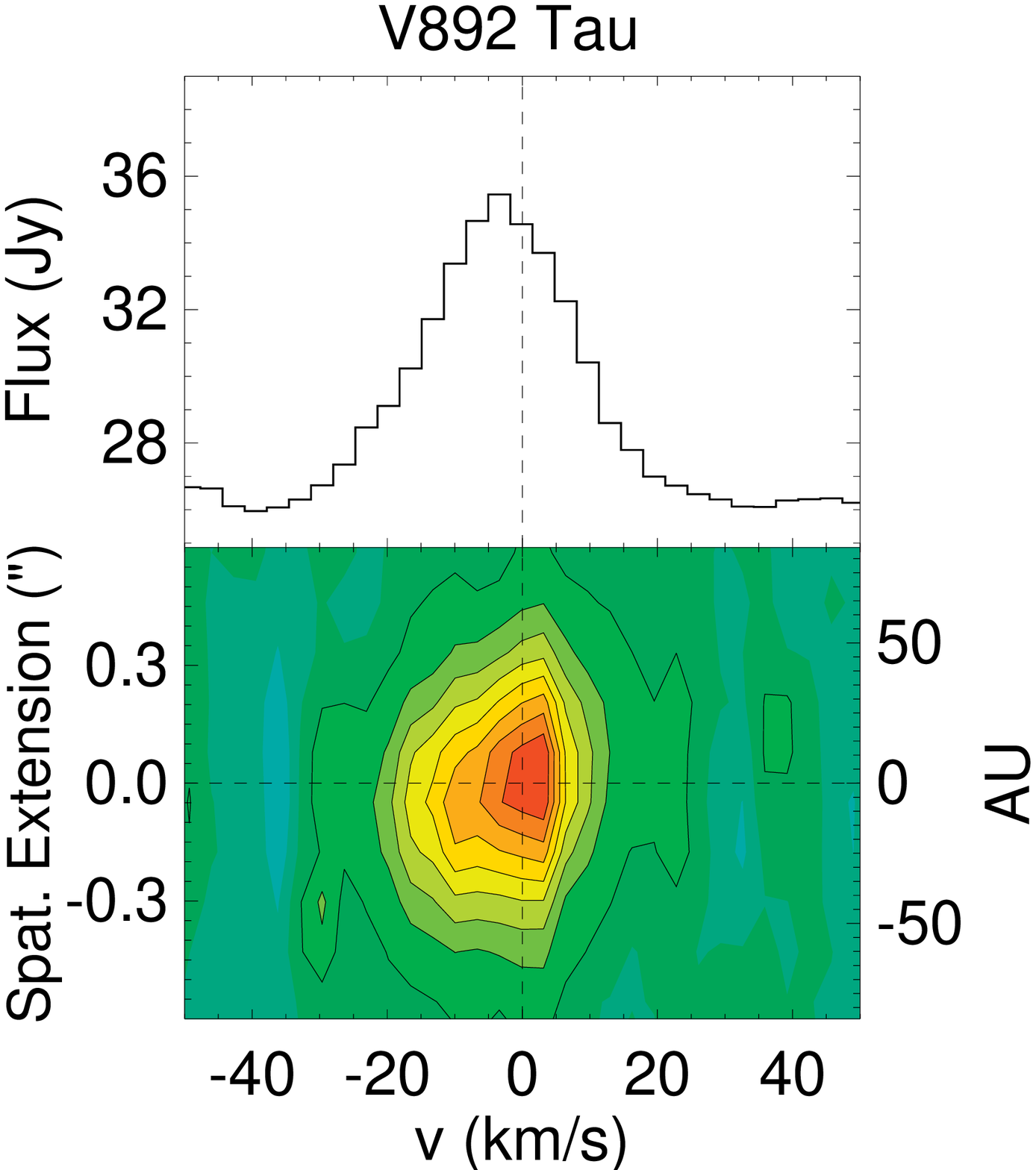}
	  		\end{array}$  
		\end{minipage}
     \caption{Upper panels: Observed VISIR spectrum of  \object{CoKu Tau 1},  \object{FS Tau A}, and  \object{V892 Tau}. Lower panels: Position velocity diagrams after subtraction of the continuum PSF. The contour levels start at $3\sigma$ and increase with a step of $5\sigma$. The PSF FWHM is $\approx 0\farcs3$.}				
     \label{fig:spat_ext}
 \end{figure*}

The spatial extent of the emission can be quantified by subtracting the average PSF observed in the continuum from the 2D spectrum, as presented in the position velocity diagrams in Figure~\ref{fig:spat_ext} (contours start at $3\sigma$ with an increase of $5\sigma$ at each step), bearing in mind the limited spatial resolution imposed by the instrumental PSF.
 \object{CoKu~Tau~1} presents symmetric emission centered at the stellar rest velocity, with no extended emission beyond  the $0\farcs3$ (40 AU at the distance of Taurus) spatial resolution of the VLT in the mid-infrared. The average PSF in the continuum is also consistent with a point source, indicating that the $0\farcs24$ companion \citep{padgett:1999aa} was not detected.
 The centroid of the emission at negative velocities is slightly displaced to ``negative'' spatial extension (about 10~AU) whereas the centroid at positive velocities goes to ``positive'' spatial extension (also about 10~AU). This could be a signature of disk rotation, but considering the fact that the slit was oriented with PA=$0^\circ$, i.e., close to the direction of the bipolar jet (PA=$210^\circ$), the centroid's displacement could also reflect emission by the bipolar jet. We address this point in the discussion (Section~\ref{discussion:model_comp}).
The emission from  \object{FS~Tau~A} is centered at negative velocities, with evidence of extension toward negative velocities, indication of photoevaporative wind. 
The symmetry of the emission is much less evident than in CoKu~Tau~1. The $3\sigma$ level extends up to velocities of $\pm 40$~\kms. The \neon\ PV diagram in  \object{FS Tau A} suggests some extended emission beyond $0\farcs3$ that cannot be attributed to the close $0\farcs24$ companion \citep{hartigan:2003aa}.
The \neon\ emission from  \object{V892~Tau} shows an extension toward the blue side of the spectrum.   Its PV diagram also indicates extended emission beyond $0\farcs3$.

\subsubsection{Comparison with {\it Spitzer} line fluxes}

We now compare the line fluxes obtained with VISIR with the fluxes obtained previously from {\it Spitzer} low-resolution spectra.
In general, VISIR fluxes tend to be lower than {\it Spitzer} fluxes (reported in Table \ref{table:fit_results} column 3).
Focusing only on the detected lines, we found that in one case (V892~Tau) the flux obtained with VISIR corresponds to the {\it Spitzer} upper limit.
The difference between \spit~and VISIR fluxes can go as high as one order of magnitude.
In the case of  \object{MHO~1/2}, the \spit~flux reported is for the unresolved binary, but the \neon~line was detected only in  \object{MHO~1} with VISIR.
We found that for  \object{FS~Tau~A} and  \object{SST042936+243555} the VISIR flux is between $30$ and $40\%$ lower than the \spit~flux. 
For  \object{CoKu~Tau~1},  \object{V853~Oph}, and  \object{EC~92} the difference between {\it Spitzer} and VISIR fluxes are high, the flux obtained with VISIR observations represent between $20-30\%$ of the {\it Spitzer} flux.
The variation between {\it Spitzer} and VISIR fluxes has been previously reported by \citet{pascucci:2011aa}, where they found on the order of $30\%$ for  \object{TW~Hya}. 
\citet{sacco12} also found evidence of the flux discrepancies in their targets for Class I and Class II stars, arguing that the \neon\ emission is extended at least in Class I stars, 
and partially in Class II stars. However, they found flux ratios between \spit\ and VISIR that are consistent within a factor of 2 for pre-transition and transition disks, suggesting that the emission is produced close to the star, within $\sim 20-40$~AU.
In our study, the reason for the observed variation  remains unclear, although plausible but unlikely explanations might be unidentified flux calibration issues, slit losses in VISIR observations due to an incorrect centering of a star in the slit, unexpected telescope drifts, or to atmospheric conditions. A faint, very broad line could also remain undetected. The most likely explanation for the flux discrepancies is the different beam sizes between {\it Spitzer} and VISIR that translate into possible extended emission unresolved and detected in {\it Spitzer} but undetected with VISIR. Indeed, \spit~spectra are extracted from a region covering the whole slit width, for IRS high-resolution this means a region of $\sim 11\arcsec$, i.e., about 1500~AU at 140~pc.

We note that  \object{L1551~IRS5} was selected for this program because a bright \neon~line was detected in the \spit~spectrum, with a flux of $5.8 \times 10^{-13}$~erg~s$^{-1}$~cm$^{-2}$, i.e., a luminosity of the order of \lne~$\sim 10^{30}$~erg~s$^{-1}$.
Although the line was not detected in spectroscopy mode, the target was detected in the [Ne~II] narrow band image, likely due to the strong continuum emission.
This Class~I star is known to drive a highly collimated bipolar outflow observed in the optical, near-infrared and radio (e.g., \citealt{davis:2003aa,pyo:2009aa,wu:2009aa}).
The position angle of the outflow is measured at $260\degr$ for the northern component and $235\degr$ for the southern component \citep{pyo:2009aa}.
The VISIR observations were performed with a P.A.$=0\degr$.
The non-detection of \neon~in the spectrum suggests that the emission originates in the outflow, at distances larger than $\approx 0\farcs4$ from the central star, although the short integration in the acquisition image did not allow the detection of extended emission. 
We did not repeat the observations at a different P.A. because the outflow explanation is the most likely in this case.

\subsection{Optical emission lines detected in UVES spectra}
\label{subsect:optical_lines}

Low-excitation forbidden lines are used as diagnostics of the physical conditions of the gas in the wind and shock environments of \tts~stars (e.g.~\citealt{hamann:1994aa,gomez-de-castro:1993aa,ouyed:1994aa}).
For example, the ratio $[\rm{S~II}] \lambda 6716/[\rm{S~II}] \lambda 6730$ is strongly dependent on the electron density $n_e$, while the ratios $[\rm{S~II}] \lambda 6730/[\rm{O~I}] \lambda 6300$ and $[\rm{N~II}] \lambda 6584/[\rm{O~I}] \lambda 6300$ are dependent on the electron temperature, $T_e$.  
The \oxy~$\lambda 6300$ line is usually observed in \tts spectra (e.g., \citealt{cabrit:1990aa}), often exhibiting a broad high velocity component blueshifted up to several hundred km~s$^{-1}$ (HVC) and a low-velocity component (LVC)  which is narrower and blueshifted by only a few km~s$^{-1}$ (see e.g., \citealt{hartigan:1995aa}).   
The LVC and HVC are thought to have different origins; the HVC is believed to be tracing a jet, while the LVC is interpreted as a tracer of a photoevaporative disk wind by \citet{ercolano:2010aa}.
In particular, these authors proposed the presence of both \neon~and \oxy~as an indicator of an X-ray photoevaporative disk wind.
Permitted emission lines of Fe~II are also observed in the optical. For example, \citet{beristain:1998aa} found line profiles showing a narrow and a broad component that were interpreted as being produced by different kinematic zones: the narrow component is produced by turbulence and broadened by stellar rotation, originating in gas located in the post-shock region, and the broad component is produced in gas infalling in the accretion funnel. 
Furthermore, He~I emission lines are presented as a powerful diagnostics, because their high excitation potential the emission is restricted to regions of high temperature or close to the ionizing source \citep{beristain:2001aa}. 

The optical spectra obtained with UVES show several of these emission lines, whose profiles can be compared with the profile of the \neon~line from VISIR to understand the circumstellar environment of the sources and provide hints on the origin of the \neon~lines observed. 
In Table~\ref{table:UVES_line_prop} we present the properties of the optical lines detected in the spectra of  \object{CoKu~Tau~1},  \object{FS~Tau~A}, and  \object{V853~Oph}.
In the second column we include the rest wavelength of the lines taken from the NIST Atomic Spectra Database \footnote{http://physics.nist.gov/}.
The line center with respect to the systemic velocity of the star, and {\rm FWHM} (Cols. 3 and 4) were obtained by fitting a Gaussian profile to the lines, and the errors reported come from the fit. The line fluxes (Col. 5) were calculated using as continuum the stellar flux derived from the $R$ magnitude and the Johnson zero point photometry.
In Col. 6 we present the equivalent widths (EW) calculated by integrating the line fluxes within $3\sigma$ of the Gaussian fit.

\addtocounter{table}{1}

\section{Discussion}
\label{discussion}

\subsection{Origin of \neon~emission: observations and models}
\label{discussion:model_comp}

The emission of \neon~from \protop~has been discussed in a series of articles that model the irradiation of the disk by high energy photons, each of them gives a different possible scenario for the formation of this emission line (see \citealt{glassgold:2007aa,alexander:2008aa,gorti:2008aa,ercolano:2010aa}).
Observations at high spectral and spatial resolution provide line profiles and shifts that can be used to test the different models.

\citet{glassgold:2007aa} predicted \neon~emission from an X-ray irradiated disk atmosphere.
In this model, the \neon~emission comes from a region within $20$~AU from the star.
The predicted line profile is double peaked with extended wings associated with material at small radii, although the authors did not give any estimates of the line widths expected.

\citet{alexander:2008aa} modeled the \neon~line profiles from a photoevaporative disk wind driven by stellar EUV photons, assuming a spectral resolution of $R\sim 30 000$.
The model included different disk inclination angles. 
The line profiles obtained varied with the disk inclination considered. 
A disk viewed face-on (inclination angle $i=0^{\circ}$) showed a single peaked line blueshifted by typically $<10$~\kms. 
On the other hand, a disk viewed edge-on ($i=90^{\circ}$) showed a double-peaked profile. 
The emission is expected to be centered at the stellar velocity with one peak blueshifted and the other redshifted. 
The line widths predicted are {\rm FWHM}~$\sim 10$~\kms~for the face-on case and {\rm FWHM}~$\sim 30$~\kms~for the edge-on case.
In addition, a disk with an inner hole of $9$~AU (in both dust and gas) is also modeled.
For the face-on case, the line width and shift are equivalent to the disk without an inner hole.
In the edge-on case, the line was narrower for the model with an inner hole than for the model without an inner hole ({\rm FWHM}~$\sim 20$~\kms). In addition, the line was  double peaked and centered at the stellar rest velocity.
A narrower line is in fact expected since the material closer to the star has larger rotational velocity and gives rise to the broadening of the line.

\citet{ercolano:2010aa} considered the photoevaporation of the disk driven by EUV and X-rays and modeled the profiles for a series of emission lines, putting special attention on \neon~and \oxy~($\lambda6300$).
The model considered a disk with and without inner holes of $9$, $14$, and 30~AU at different inclination angles between $0^{\circ}$ and $90^\circ$. Furthermore, the X-ray luminosity was also variable between $\log (L_{\rm X})=28.3$ to $\log (L_{\rm X})=30.3$.
The profile of the \neon~line for the model without an inner hole varied according to the disk inclination.
A face-on disk showed a narrow profile, with a {\rm FWHM} between $\sim 4$~\kms~for low \lx~and $\sim 10$~\kms~for high \lx.
For the edge-on case, the line profiles are larger but do not change much with \lx; the {\rm FWHM} are between $\sim 20$~\kms~and $\sim 23$~\kms~for the low and high \lx, respectively.
Only in the case of a face-on disk and low \lx~was the line found to be centered at the stellar velocity; in all other cases the line was blueshifted by up to $\sim 5$~\kms. 
All the profiles obtained were asymmetric, with a shoulder toward the blueshifted part, which became more evident for the high \lx~model.

Using the predictions of such models we can attempt to give an interpretation to the line profiles observed.
Following the order of the panels of Fig.~\ref{fig:visir_ne2_spect} we can identify three regimes for the \neon\ emission: 

{\it Highly blueshifted \neon~emission from jets}: The line profiles shown in the upper panel of Fig.~\ref{fig:visir_ne2_spect} correspond to this group and they are characterized by a large velocity shift toward the blue. The emission is likely to originate mainly from shocks in jets.
For  \object{MHO-1} and  \object{SST~042936+243555} there is no available measurement of the radial velocity (we assumed the average radial velocity of the Taurus Molecular Cloud) but since they are observed with high velocity shifts, $-122$ and $-89$~\kms~respectively, we conclude that the emission is likely originating in a jet.
 \object{V853~Oph} and  \object{EC~92} also likely belong to this group, with velocity shifts of $-36$ and $-31$~\kms~respectively, despite the low peak velocities, perhaps due to inclination effects.

There is no record in the literature of jets detected in our sample of stars with highly blueshifted \neon~emission. 
In the case of  \object{SST~042936+243555}, this is likely due to the low number of observations dedicated to look for jet or outflow emission published.
From a search in the literature,  spectroscopic observations in the optical or in the near-infrared do not show any evidence of outflow or jet emission in  \object{MHO-1},  \object{V853~Oph}, or  \object{EC~92}.

\citet{shang:2010aa} modeled the profiles of \neon~and \oxy~($\lambda 6300$) expected from a jet.
The centroid of the lines show a large shift toward the blue, reaching $-200$~\kms, for an inclination angle of $45^{\circ}$.
The lines are asymmetric with a broad shoulder toward the red side of the spectrum. 
The excess emission toward the red can reach up to $+100$~\kms, depending on the inclination angle.
The profiles of the model were calculated for a resolution of $1$~\kms, much higher than the resolution achieved with our VISIR observations. 
However, the dominant feature is a large blue-shifted peak, like observed in  \object{MHO-1},  \object{SST~042936+243555},  \object{V853~Oph}, and  \object{EC~92}.

{\it Weakly blueshifted \neon~emission from photoevaporative wind:} The centroids of the lines in this category are close to the stellar rest velocity or show a small velocity shift. 
To this group belong  \object{FS~Tau~A} and  \object{V892~Tau}.
In particular, the \neon~profile from  \object{V892~Tau} shows an excess toward the blue, and a line width of $25$~\kms.
The star is known to be a binary of separation $\sim 5$~AU. 
The circumbinary disk was detected by \citet{monnier:2008aa} through infrared imaging.
The inclination of the disk is estimated to be ${i=60\degr}$ and the disk inner hole to be $\sim 35$~AU. 
  \object{FS~Tau~A} is also a close binary system, with separation $0\farcs24$ (\citealt{hartigan:2003aa}, equivalent to $34$~AU at the distance of Taurus) and surrounded by a circumbinary disk that extends up to 630~AU \citep{hioki:2011aa}.
The spatial extension of the emission (Figure~\ref{fig:spat_ext}) in both systems shows an excess toward the blue side of the spectrum, i.e., an indication of a photoevaporative disk wind.

 {\it The special case of  \object{CoKu~Tau~1}:}
 The emission line detected in  \object{CoKu~Tau~1} shows a flat-topped, broad profile with a FWHM of 55~\kms, and wings extending up to velocities close to $\pm 80$~\kms. The emission is also spatially unresolved. The immediate interpretation is that the \neon\  line originates from the disk atmosphere, from material located at small disk radii. 
 However, the star is known to be viewed close to edge-on (${i=87^\circ}$, \citealt{robitaille:2007aa}) and is known to have a bipolar jet detected at both positive and negative velocities relative to the star \citep{movsesyan:1989aa,eisloffel:1998aa}. 
  Photoevaporation models at nearly edge-on inclinations predict large wings, but due to model construction they never extend beyond $\pm 40$~\kms~(e.g., \citealt{alexander:2008aa,ercolano:2010aa}). Perhaps a fast photoevaporative wind could explain the broader wings observed in  \object{CoKu~Tau~1}. \citet{bast:2011aa} argued also that broad centrally peaked CO line profiles in protoplanetary disks could not be explained by inclined disks in Keplerian rotation, but that a combination of emission from the inner part of the disk and a slow moving disk wind could explain the line profiles. Although the disk interpretation for  \object{CoKu~Tau~1} is more physically interesting, the bipolar jet explanation is as valid: first, its \neon\ flux is rather high with \spit\  and VISIR and is more typical of fluxes detected from jets. Second, although the PV diagram shows spatially unresolved emission, a careful analysis indicates peaks at ``negative'' spatial extension for negative velocities, and peaks at ``positive'' spatial extension for positive velocities. Considering the fact that we used a NS slit orientation with VISIR (and UVES) while the jet is oriented with PA=$210^\circ$ with positive velocities toward the South (and inversely for the counterjet; \citealt{eisloffel:1998aa}), the centroids of the \neon\  peaks are consistent with the bipolar jet interpretation. On the other hand the direction of rotation of the disk in  \object{CoKu~Tau~1} is unknown, and the \neon\ peaks could also trace this rotation. In fact, the peaks extend up to $\pm 10$~AU. In conclusion, the origin of \neon\ emission in  \object{CoKu~Tau~1} is difficult to ascertain and could be either from the disk atmosphere or from the bipolar jet, with a possible contribution by a photoevaporative wind. A comparison with the optical forbidden line (see below) does not allow us to constrain better the origin of the \neon\  line.

\begin{figure*}
\begin{center}
\includegraphics[width=1.\textwidth]{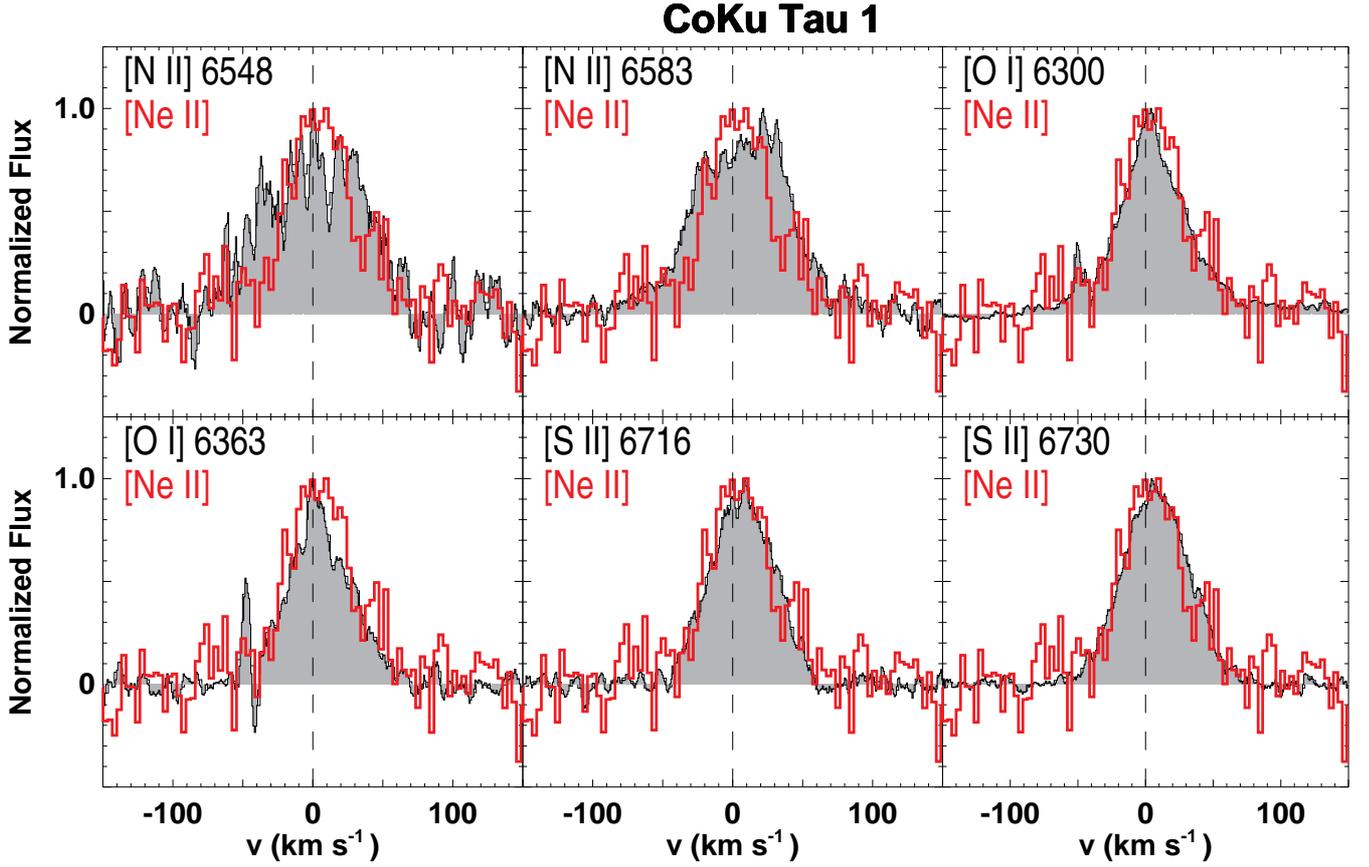}
\end{center}
\caption[Lines detected in the UVES spectrum of CoKu Tau 1]{Lines detected in the UVES spectrum of  \object{CoKu Tau 1} plotted as a filled grey area compared to the profile of \neon~(red line). The spectra were continuum subtracted and normalized to the peak of the lines to allow comparison.}
\label{fig:CoKU_optical_neon}
\end{figure*}

\begin{figure*}
\begin{center}
\includegraphics[width=1.\textwidth]{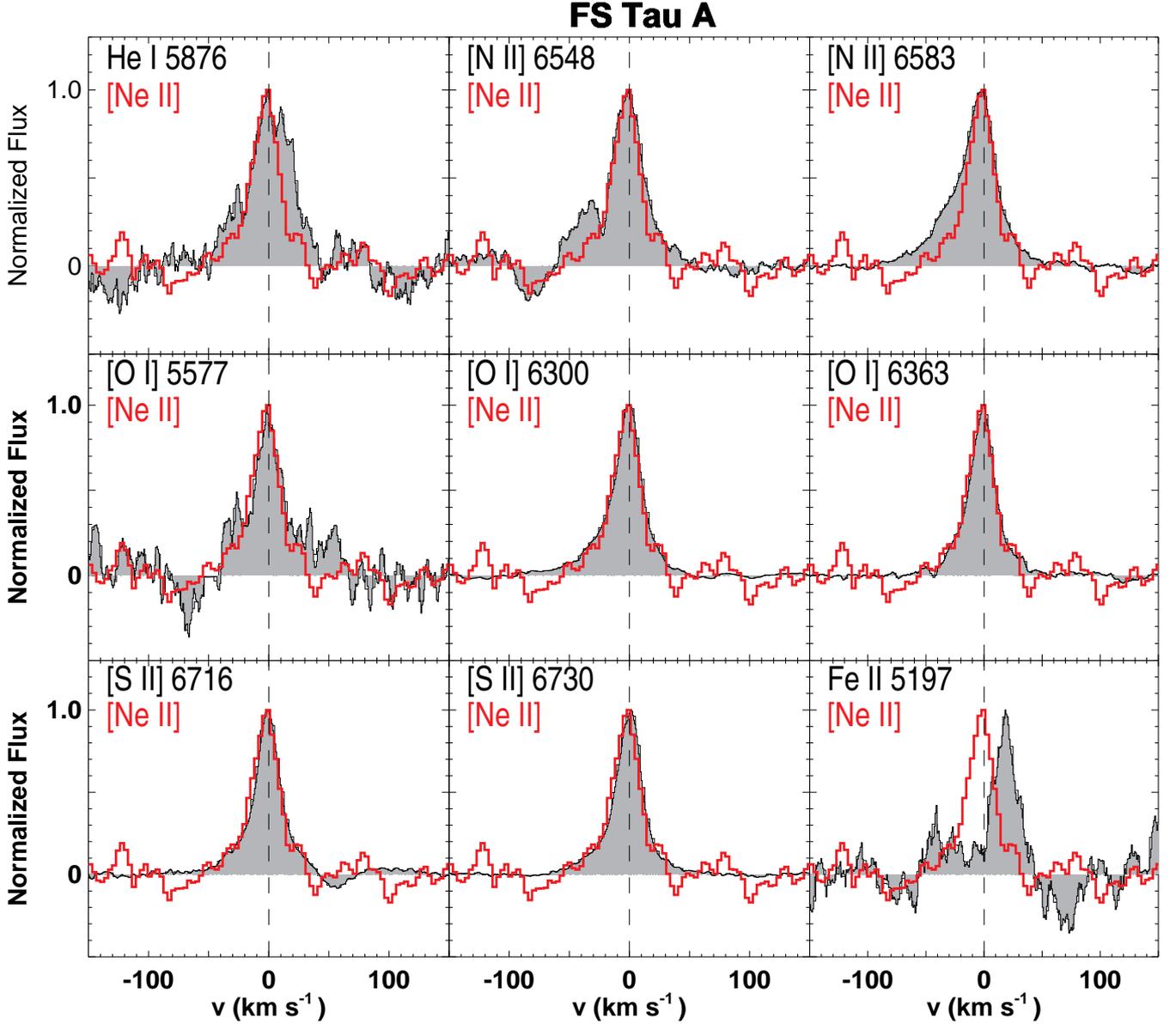}
\end{center}
\caption[Lines detected in the UVES spectrum of FS Tau A]{Profile of the optical lines detected with UVES and the profile of the \neon~line for  \object{FS~Tau~A}. The color code used is the same as Fig.~\ref{fig:CoKU_optical_neon}.}
\label{fig:FSTAU_optical_neon}
\end{figure*}

\begin{figure*}
\begin{center}
\includegraphics[width=1.\textwidth]{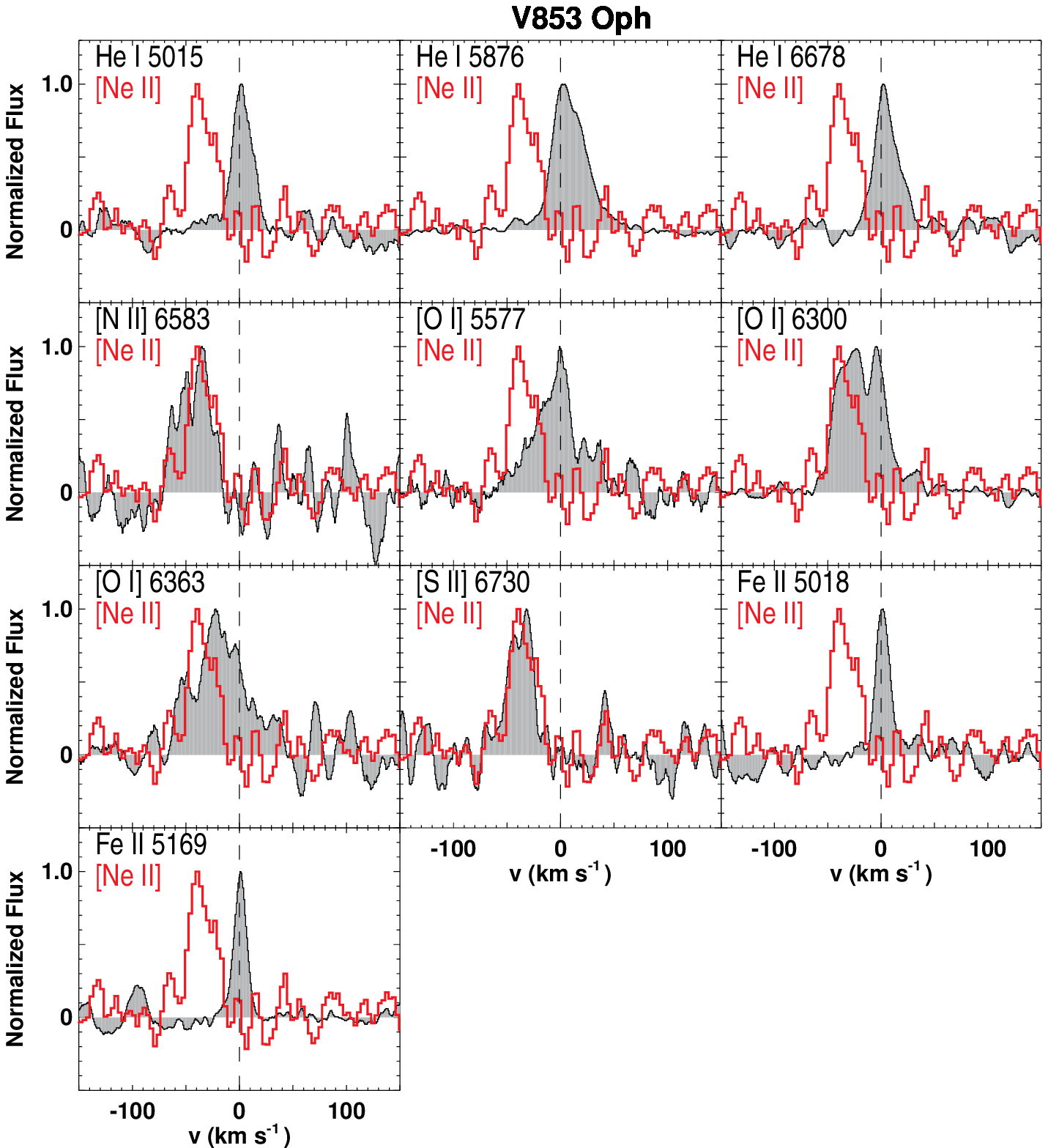}
\end{center}
\caption{Profile of the optical lines detected with UVES and the profile of the \neon~line in  \object{V853~Oph}. The color code used is the same as Fig.~\ref{fig:CoKU_optical_neon}.}
\label{fig:V853OPH_optical_neon}
\end{figure*}

\subsection{Comparison between \neon~profile and optical lines}

We can further compare the profiles of the \neon~line with the profiles of the optical lines for the three targets observed in our UVES program.
Figures \ref{fig:CoKU_optical_neon}, \ref{fig:FSTAU_optical_neon}, and \ref{fig:V853OPH_optical_neon} show in black line and grey area the profile of the \neon~line and in red the profile of the optical lines detected.
The spectra were continuum subtracted and normalized to the peak of the line to allow comparison.

\begin{itemize}
\item{ \object{CoKu~Tau~1}}
\end{itemize}

The profiles of the optical forbidden lines are generally compatible with the profile of the \neon~line, i.e., similar center and {\rm FWHM}. The [N~II] lines are slightly broader than \neon, in contrast to the [O~I] and [S~II] lines which are of similar width (or slightly smaller in the case of oxygen). The H$\alpha$ line is much broader, possibly tracing additional regions than the optical forbidden lines. The Fe~II line ($\lambda 6432$) is also much broader (FWHM~$\sim 250$~\kms). The optical forbidden lines are only shown in Figure \ref{fig:CoKU_optical_neon}, whereas the Fe~II and H$\alpha$ lines appear in the Appendix (Fig.~\ref{fig:CoKU_optical_lines}). The shapes of the two detected [N~II] lines  ($\lambda 6548$ and $\lambda 6583$) are more difficult to interpret, they appear double or multiply peaked but this might also be due to a lower signal-to-noise ratio in that part of the spectrum. We note that the optical line profiles are compatible with those of \citet{white:2004aa} from HIRES data. The optical emission spectrum is completely dominated by optical forbidden lines (and H$\alpha$), which is typically seen in edge-on Class I stars \citep{white:2004aa}. The origin of these optical lines is unclear but could either be due to the disk atmosphere or from the inner bipolar jet. The profile of the \neon\  line is  consistent with the forbidden lines (Fig.~\ref{fig:spat_ext}), which suggests that the line may either originate from the same region. In the bipolar jet interpretation, the broader [N~II] lines may include a faster component, possibly because they would form closer to the faster jet axis, whereas the [O~I], [S~II], and [Ne II] lines form closer to the slower jet surface. The comparison of the optical forbidden lines and the \neon\  line does not allow to determine conclusively the origin of the lines, although the bipolar jet origin is the most likely one.

\begin{itemize}
\item{ \object{FS~Tau~A}}
\end{itemize}

The profile of the \neon~line in FS~Tau~A is generally similar to the profiles of the optical forbidden lines, and to He~I ($\lambda 5876$; the line is, however, broader) and Fe~II ($\lambda 5197$; the line is redshifted with respect to \neon\  and the stellar radial velocity). 
The optical forbidden lines are nearly centered at the stellar radial velocity or slightly blueshifted with a {\rm FWHM}~$\sim 30$~\kms.
The profiles are consistent with a photoevaporative wind flow, like the ones modeled by \citet{ercolano:2010aa}, nearly centered with a shoulder toward blueshifted velocities.
This blueshifted shoulder is larger for [N II] ($\lambda6583$). 
The observed line profiles are consistent with the modeled profiles considering the disk inclination, estimated at ${i=30-40^\circ}$ \citep{hioki:2011aa}. 
Furthermore, the presence of \neon~and \oxy~are indications of an X-ray driven photoevaporative flow, according to \citet{ercolano:2010aa}.
In fact, the low-velocity component of the \oxy~line can only be reproduced in the presence of X-rays.
A comparison with the profile modeled by \citet{alexander:2008aa} is more difficult since only the profiles for the edge-on and face-on disks are shown.
Nevertheless the {\rm FWHM} for the case $i=45^\circ$ is reported and is compatible with the value derived from the observations: {\rm FWHM$_{\rm model}$}$= 28$~\kms~versus {\rm FWHM$_{\rm obs}$}$= 27$~\kms.

\begin{itemize}
\item{ \object{V853~Oph}}
\end{itemize}

V853~Oph shows a \neon~line blueshifted by $36$~\kms, with a {\rm FWHM}~$=26$~\kms.
The large velocity shift observed suggest that the line is emitted by a jet.
In this star, the profiles of the optical lines are not always compatible with the profile of the \neon~line; we can classify them according to their shifts and widths:

{\it Lines emitted by the star or the disk}: These are He~I ($\lambda 5015, 5876, 6678$) and Fe II ($\lambda 5018, 5169$). They show different shapes; the He~I lines are asymmetric with an excess toward redshifted emission, while the Fe~II lines are well centered and narrow. 

{\it Lines compatible with photoevaporation}: These are [O~I] ($\lambda 5577, 6300, 6363$).
The \oxy~lines show different shapes; the $\lambda 5577$ line is fairly well centered at the stellar rest velocity and shows the blue shoulder typical of photoevaporative wind flows, and is also compatible with the disk inclination angle (${i=32^\circ}$, \citealt{andrews:2010aa}).
The $\lambda 6300$ line, on the other hand, shows two velocity components: one  close to the stellar rest velocity and the other with a higher velocity shift toward the blue, probably due to a jet. The separation between the components are believed to be real, although there is likely a moderate absorption between the components due to  water molecular absorption. Indeed, 
although oxygen emission lines are present in the sky spectrum, they are removed in the data reduction process. But  molecular absorption lines from the atmosphere can only be corrected with observations of a telluric standard. Since we do not have a telluric calibrator for the optical spectra, we used archival spectra of early type stars to look for contamination of atmospheric absorption lines in the profile of \oxy. There is, indeed, a 20-30~\% absorption feature at the proper position, suggesting that some but not all of the absorption feature observed  in the $\lambda 6300$ line is due to a water absorption line.

{\it Lines emitted in a jet}: these are the [N~II] ($\lambda 6583$)  and the [S~II] lines, and the high-velocity component of the [O~I] $\lambda  6300$ line.
The [S~II] ($\lambda 6730$) line matches the \neon~profile, with similar center and {\rm FWHM}, whereas the [N~II] line is slightly broader and shows emission at higher speeds, but the signal-to-noise ratio is not large. 
The line profiles differ from the profiles predicted by \citet{shang:2010aa}; they are narrow and the peak is located at velocities lower than predicted by the model; $-37$~\kms~against $\sim 100-200$~\kms.

In conclusion, at least in the case of V853 Oph, the detected \neon\  line is not a tracer of photoevaporative wind since it traces a high-velocity component of a jet. Possibly the \neon\  component from a photoevaporative wind exists, but it is much fainter than the jet component detected with \spit\ and VISIR.

\subsection{Summary of \neon~ground based observations.}
\label{subs:discussion_sample}

Putting together previous results (\citealt{herczeg:2007aa,van-boekel:2009aa,najita:2009aa,pascucci:2009aa,pascucci:2011aa,sacco12}) with the ones from our program, there is a sample of 26 objects, for which \neon~emission is detected from the ground with high spectral resolution to derive the kinematics of the \neon~emission. The study by \citet{sacco12}, published while this paper was being refereed, provided 12 new detections among 32 targets and was able to provide some conclusions about the origin of \neon\  emission, in particular for Class I sources and for transition and pre-transition disk sources. Our study focusses more on Class II stars and provides additional comparison of the \neon\  line emission with high-resolution optical lines, in particular forbidden lines such as [O~I], [S~II].
We can attempt to extract some general conclusions from the results we have in hand up to now, keeping in mind that  ground based observations are challenging in terms of sensitivity.
Studies based on \spit~observations showed a weak correlation between the \neon~luminosity and the X-ray luminosity, in particular for optically thick disks sources.
This correlation is expected if X-rays are responsible for the neon emission. \citet{sacco12} found no obvious correlation between the \neon\ and X-ray luminosities in their sample. Restricting to all detected sources in which ground based observations give a line consistent with disk emission and the ones consistent with photoevaporation, there is still no clear evidence of a trend. But, 
if dropping the 5 targets in \citet{sacco12} with an observed blueshift less than 18~\kms\  and whose disk inclinations are unknown, there is a possible trend for higher neon luminosities with higher X-ray luminosities. However, we emphasize that the sample is relatively small, and there are many uncertainties (classification, inclination, variability, etc).

Another lesson from \spit~observations is that bright neon emission is expected in jet sources, and that whenever jets are present, they dominate over the disk emission \citep{gudel:2010aa,baldovin-saavedra:2011aa}.
Ground-based observations have shown that jets are significant emitters of \neon\  line emission (e.g., \citealt{van-boekel:2009aa}, \citealt{pascucci:2009aa}, \citealt{sacco12}, this work). The detected lines in jet sources tend to be fainter than lines compatible with disk or slow wind emission (see also Fig.~4 in \citealt{sacco12}).
This can be explained because the larger beam of \spit~was able to capture most of the extended emission produced by shocks, but this emission is filtered out by the narrow slit used in ground based observations. Jet activity is closely related to the evolution of pre-main sequence stars and to their accretion history. Outflows and jets are more powerful during the earlier stages and they tend to decrease in intensity as the star evolves toward a state of lower level of accretion.
Unfortunately, there is little information in the literature about the stars of the sample showing neon emission consistent with outflows/jets.
They are generally classified as Class~II sources, and the levels of disk mass accretion rate are in the typical level of T Tauri stars ($\sim 10^{-8}$~\msun~yr$^{-1}$; e.g., \citealt{gudel:2007aa,gudel:2010aa}, and references therein).

 \citet{sacco12} also argue that the ratio between \neon\  fluxes measured by \spit\ and VISIR are consistent for transition disk and pre-transition disk sources. Our detected sample does not include targets with such a classification, but Class II stars (except V892 Tau) that show VISIR fluxes several times lower than \spit\ fluxes, consistently with the Class II stars observed by \citet{sacco12}. The latter argued that the emission in transition and pre-transition disk objects would be located within a few AU from the central star, in contrast with Class II sources that would emit \neon\  from both the inner region and an extended envelope. In our three targets with low velocity shifts, extended emission is found for both  \object{FS Tau A} and  \object{V892 Tau}, whereas  \object{CoKu Tau 1} shows no evidence of extended emission (but its emission is either due to a bipolar jet or a disk). Thus, overall, our data support the interpretation by \citet{sacco12} that extended emission is present in some Class II sources.

In three stars the \neon~line could be interpreted with disk emission: CoKu Tau 1 (this work), AA~Tau, and GM Aur \citep{najita:2009aa}.
The first two are classical optically thick disks and GM~Aur is a more evolved transitional disk. The first two also have high inclination angles:
 $i= 87^\circ$ ( \object{CoKu~Tau~1}; \citealt{robitaille:2007aa}) and $i=75^\circ$ (AA Tau, \citealt{bouvier:1999aa}), where as  \object{GM Aur} has $i=54^\circ$ \citep{simon:2000aa}.
 As discussed in this paper, the disk origin for CoKu~Tau~1 is difficult to ascertain and could be attributed to the bipolar jet. Interestingly, \citet{cox:2005aa} reports
 the presence of a jet in  \object{AA~Tau}. Although \citet{najita:2009aa} interprets the broad (FWHM=70~\kms) \neon\  line as due to the disk, a similar situation
 as in CoKu~Tau~1 could occur, and the emission could be due to a bipolar jet in AA~Tau. However, in the case of GM Aur, its moderate inclination, its transitional disk status, and the absence of a reported jet all suggest that the emission may really come from the disk. In any case, the X-ray luminosities of the disk sources are of the order of $10^{30}$~erg~s$^{-1}$, sufficient to effectively heat the disk and to be responsible for the neon emission.

Different studies suggest that soft X-rays can drive powerful photoevaporative winds that can disperse the disk effectively. 
Observationally, these winds translate in emission lines showing a small blueshift. 
These winds are stronger, and are effective once the mass accretion rate has fallen to values lower than $10^{-8}$~\msun~yr$^{-1}$ for an X-ray luminosity  \lx$~\sim 2~\times 10^{30}$~erg~s$^{-1}$. Out of 12 targets from \citet{sacco12}, ten show evidence of small blueshifts within 18~\kms of the stellar velocity. Only two targets were classified as Class II sources, while the rest was classified as transition or pre-transition disk sources. In our sample, FS Tau A and V892 Tau also show a blueshift, and three additional sources show such blueshifts in \citet{pascucci:2009aa}.
Interestingly, \citet{sacco12} found evidence that the FWHM of the detected lines in their sample increased 
with increasing blueshift. Our sample does not show this trend, with the majority of lines with FWHM around 26-27~\kms, except for CoKu Tau 1 that is much broader and EC 92 which is narrower but slightly above the VISIR instrumental resolution of 10~\kms. In fact, a plot of FWHM vs $v_{\rm peak}$ from both our and Sacco's data sets shows no evidence of a correlation. The high number of \neon\  lines consistent with photoevaporative winds by \citet{sacco12} could suggest that photoevaporation of the disk is an effective mechanism. However, the significant number of non-detections of \neon\  in our study may also indicate that photoevaporative winds may not be as efficient as suggested in all young disk-bearing stars. In addition, our detected sources show different behaviors, from jets, photoevaporative winds, to disk emission. It may, nevertheless, be  that photoevaporation is the most efficient mechanism for \neon\  emission in transition and pre-transition disk sources, as argued by \citet{sacco12}. Disk emission appears, on the other hand, rare and subject to uncertainties such as the disk inclination.

\section{Summary and Conclusions}
\label{conclusions}

We presented observations of the \neon~line at $12.81$~\mic~at high spectral and spatial resolution obtained with VISIR-VLT in a sample of selected pre-main sequence stars  with previously detected neon emission with \spit. 
The emission line was detected and spectrally resolved in seven stars. 
We interpreted the profiles and shifts according to three emitting mechanisms:

\begin{itemize}
\item Lines consistent with photoevaporation. 
These are the lines detected in FS Tau A and the Herbig Be star  \object{V892 Tau}. They show small shifts with respect to the stellar velocity and  FWHM $ \sim 26$~\kms. 

\item Lines consistent with shocked material in outflows/jet. The line centroids are blueshifted at large velocities in four stars:  \object{MHO-1},  \object{SST042936+243555},  \object{V853 Oph}, and  \object{EC~92}. To our knowledge, there is no previous record in the literature of the detection of a jet in these stars.
On the other hand, for several stars known to be jet driving sources we did not detect neon emission.
It is possible that in these cases the orientation and width of the slit used prevented us from detecting neon emission at large velocities.

\item Lines consistent with disk emission. This is possibly the case of  \object{CoKu Tau 1} since the line is centered at the stellar velocity, has a symmetric profile and large FWHM $\sim 55$~\kms. The line also presents wings extending toward velocities close to $100$~\kms. The emission of the fast rotating gas is likely coming from regions close to the star. However, an interpretation of the line due to the bipolar jet cannot be excluded and may, in fact, be better in view of the high \neon\  luminosity in  \object{CoKu~Tau~1}.

\end{itemize}

For three stars in the sample, optical observations with UVES allowed the determination of the radial velocity, a crucial measurement for determining the line shifts.
The profiles of the forbidden lines detected in the optical spectra were compared with the profile of the \neon~line to better constrain the emission mechanism.
In general we found a good agreement between the profiles of the optical forbidden lines and the \neon~lines, indicating a common emitting mechanism.
Studies that combine observations in the two bands need to be done in a more systematic way.

Combining our results with previous studies based on high-spectral and spatial resolution of neon, we attempt to extract more general conclusions on the emitting mechanism, keeping in mind that the sample is incomplete and biased toward large neon luminosities.
For stars with neon emission compatible with disk emission and photoevaporative winds, a relation between \neon\ and X-ray luminosities is difficult to see. \citet{sacco12} found a correlation between the \neon\ line FWHM and the peak velocity, but our study does not confirm this trend. In fact a combination of \citeauthor{sacco12}'s and our data sets shows no trend.

The mass accretion rates of the stars showing lines consistent with photoevaporation are in the range at which this mechanism clears out the disk. Many stars in this group are classified as transitional disks \citet{pascucci:2009aa,sacco12}, but two stars with neon emission compatible with photoevaporation presented here do not belong to this class ( \object{V892 Tau} and  \object{FS Tau A}). These two objects are close binaries with separations of 5 and 36 AU for  \object{V892 Tau} and  \object{FS~Tau~A} respectively, and surrounded by circumbinary disks. The systems were not separated by VISIR observations. An interesting result from our study is the low detection rate of \neon~lines consistent with photoevaporation, in contrast to the study by \citet{sacco12}. Even more interestingly,  \object{CoKu Tau 1} shows evidence of broad line in \neon\  and in other optical forbidden lines, but the origin of the emission is likely from its bipolar jet, although disk emission cannot be excluded, nor a contribution from a photoevaporative wind.

A deeper knowledge of the stars studied is needed.
As an example, for the sources with neon emission compatible with jet emission, very little information is published. 
Important parameters such as mass accretion rate, or even the presence of outflow/jet activity from other tracers is not available.
Observations of the \neon~line can be used to study jets, photoevaporation, and disk emission. A larger sample is needed in order to identify the preferred emitting mechanism for the different types of stars. 
Although ground-based observations of \neon~from young stars are challenging, our work demonstrates that they are crucial to constrain the kinematics and physical region of the emitting source, and, thus, to determine the emitting mechanism of the \neon\ line.


\begin{acknowledgements}
This research has made use of the SIMBAD database, operated at CDS, Strasbourg, France.
We thank the referee, Dr. G. Herczeg, for thoughtful, and detailed comments that improved the manuscript.
C.~B-S., M.~A., and A.~C. acknowledge support from the Swiss National Science Foundation (grants PP002-110504 and PP00P2-130188).
The authors thank A. Mueller for providing the authors high-resolution synthetic spectra of low-mass stars, and ESO staff for performing UVES observations in service mode.
\end{acknowledgements}

\bibliographystyle{aa}
\bibliography{8329}

\longtab{5}{
 \begin{longtable}{lccccr}
\caption[bb]{Properties of the optical UVES lines for CoKu Tau 1, FS Tau A, and V853 Oph. The \oxy~lines at $\lambda 6300$ and $\lambda 6363$ of V853~Oph were fitted using the sum of two Gaussians to account for the low and high-velocity components (LVC and HVC), while flux and equivalent width were calculated for the sum of the two components.}\\
Line	& Wavelength &Center& \rm FWHM & Flux ($\times 10^{-15}$) 	&EW	\\
&(\AA)	&	(km~s$^{-1}$)	&	(km~s$^{-1}$) &	(erg~s$^{-1}$cm$^{-2}$)	&	(\AA)	\\
\hline
\endhead
\multicolumn{6}{c}{CoKu~Tau~1}\\
 \hline
H$\alpha$       &       $6562.819$      &       $10.2 (0.8)$     &       $103.5 (0.8)$   &       $5.86~(0.1)$    &       $-40.0$ \\
He I                    &       $5015.678$      & ... & ... &$<1.0$ &   $>-14.0$\\
He I                    &       $5875.621$      & ... & ... & $<0.01$ & $>-0.1$\\
He I                    &       $6678.151$      & ... & ... & $<0.01$ & $>-0.1$ \\
Li I                    &       $6707.760$      &       $5.1 (1.8)$     &       $35.3  ( 3.8)$  &       $0.10~(0.1)$    &       $0.7$   \\
$[$N II$]$      &       $6548.050$      &       $1.3  ( 0.8)$  &       $85.4  ( 3.9)$  &       $0.41~( 0.1)    $&      $-2.8$  \\
$[$N II$]$      &       $6583.450$      &       $8.5  ( 0.8)$   &       $77.3  ( 1.6)$  &       $1.33~(0.1)$&   $-9.1$  \\
\oxy            &       $5577.339$      & ... & ... & $< 0.06$ & $>-0.1$                \\
\oxy    &       $6300.304$      &       $4.6  ( 0.8)$   &       $58.2  ( 0.9)$  &       $ 3.21~(0.1) $  &       $-21.9$ \\
\oxy    &       $6363.777$      &       $3.9  ( 0.8)$   &       $54.6  ( 1.6)$  &       $ 1.02~(0.1) $  &       $-7.2$  \\
$[$S II$]$      &       $6716.440$      &       $7.0  ( 0.8)$   &       $53.5  ( 0.8)$  &       $0.85~(0.1      )$&     $-5.7$  \\
$[$S II$]$      &       $6730.815$      &       $7.6  ( 0.8)$   &       $57.4  ( 0.6)$  &       $2.01~(0.1)$&   $-13.7$ \\
Fe II   &       $5018.434$      & ... & ... &$<0.1$ &$>-2.0$    \\
Fe II   &       $5169.030$      & ... & ... &$<0.05$ &$>-0.7$\\
Fe II   &       $5197.577$      & ... & ... &$<0.07$ &$>-1.0$   \\
Fe II   &       $6432.680$      &       $38.1  ( 4.4)$  &       $292.8  ( 12.3)$        &       $0.9(0.1)$  &   $-5.3$  \\
\hline
\multicolumn{6}{c}{FS~Tau~A}\\
 \hline			
H$\alpha$	&	$6562.819$	&	$-6.0  ( 0.4)$	&	$71.8   ( 1.3)$	&	$832  (	5)$	&	$-16.1$	\\ 
He I			&	$5015.678$	&	... & ... &$<1.4$ & $> -0.2$ \\
He I	    		&	$5875.621$	&	$1.9  ( 0.7)$	&	$41.2  ( 1.9)$	&	$3.2  (	0.6)	$&	$-0.5$\\
He I			&	$6678.151$	&	... & ... & $<129$ & $> -2.5$\\
Li I			&	$6707.760$	&	$4.9  ( 0.2)$	&	$30.8  ( 0.4)$	&	$20.7 (	5.2)	$&	$0.4$	\\
$[$N II$]$	&	$6548.050$	&	$-1.2  ( 0.2)$	&	$30.6  ( 0.6)$	&	$47.2   (	5.2)	$&	$-1.0$	\\
$[$N II$]$	&	$6583.450$	&	$-3.4  ( 0.2)$	&	$31.8  ( 0.4)$	&	$181  (	5)	$&	$-3.5$	\\
\oxy     	&	$5577.339$	&	$-0.6  ( 0.8)$	&	$29.1  ( 2.0)$	&	$3.8  (	0.6)	$&	$-0.6$	\\
\oxy     	&	$6300.304$	&	$-0.6  ( 0.2)$	&	$26.0  ( 0.4)$	&	$295  (	5)	$&	$-5.7$	\\
\oxy     	&	$6363.777$	&	$-1.3  ( 0.2)$	&	$25.1  ( 0.5)$	&	$101  (	5)	$&	$-2.0$	\\
$[$S II$]$	&	$6716.440$	&	$-0.9  ( 0.2)$	&	$26.4  ( 0.5)$	&	$97.9  (	5.2)	$&	$-1.7$	\\
$[$S II$]$	&	$6730.815$	&	$-0.0  ( 0.1)$	&	$24.6  ( 0.3)$	&	$181  (	5)$&	$-3.5$	\\
Fe II	&	$5018.434$	&   ... & ... & $<1.0$ & $>-0.2$		\\
Fe II	&	$5169.030$	&	... & ... & $<0.2$ & $>-0.03$\\
Fe II	&	$5197.577$	&	$18.2  ( 0.6)$	&	$25.2  ( 1.7)$	&	$3.2  (	0.6)$	&	$-0.5$	\\
Fe II	&	$6432.680$	&	... & ... &$<2.1$  & $>-0.04$	\\
\hline
\multicolumn{6}{c}{V853 Oph}\\
 \hline
H$\alpha$	&	$6562.819$	&	$1.5  ( 0.3)$	&	$117.5  ( 0.8)$	&	$235.2(0.8)$	&	$-30.3$	\\
He I	 &	$5015.678$	&	$1.8  ( 0.3)$	&	$21.4  ( 0.7)$	&	$6.4  (	1.4)$	&	$-0.5$\\
He I	 &	$5875.621$	&	$3.0  ( 0.1)$	&	$36.1  ( 0.5)$	&	$35.0  ( 1.4)	$&	$-2.6	$\\
He I	 &	$6678.151$	&	$2.8  ( 0.2)$ &	$17.8  ( 0.9)$	&	$ 7.6  (0.8) $	&	$-1.0	$\\
Li I	 &	$6707.760$	&	$5.6  ( 0.2) $	&	$22.0  ( 1.6)$	&	$3.1  (	0.8)$ 	&	$0.4 $	\\
$[$N II$]$	&	$6548.050$	&  ...	& ... & $< 0.1$ & $>-0.02$ 	\\
$[$N II$]$	&	$6583.450$	& $-41.7   ( 0.8)$	& $34.3  ( 1.9)$ &	$< 1.7$ & $>-0.2$\\
\oxy			&	$5577.339$	& $-4.0  ( 0.8)$	&	$49.6  ( 1.9)$	&	$6.5  (	1.4)	$&	$-0.5$	\\
\oxy	~(HVC)	&	\multirow{2}{*}{$6300.304$}	& $-32.1  ( 0.1)$	&	$32.8  ( 0.1)$	&	\multirow{2}{*}{$ 13.2  (0.8)	$}&	\multirow{2}{*}{$-1.7$}\\
\oxy~(LVC)   &							& $-2.8  ( 0.1)$	&	$18.6  ( 0.1)$	\\ 
\oxy	~(HVC)	&	\multirow{2}{*}{$6363.777$}	& $-24.0  ( 0.1)$ 	& $26.9  (0.1)$	&	\multirow{2}{*}{$ 3.9   (0.8) $}	&	\multirow{2}{*}{$-0.5$}	\\
\oxy	~(LVC)	&		&    $-3.9  ( 0.1)$ 	& $11.2  (0.1)$	\\
$[$S II$]$	&	$6716.440$	& ...	& ... & $< 1.6$ & $> -0.02$	\\
$[$S II$]$	&	$6730.815$	&	$-37.1  ( 0.5)$	& $26.0  ( 1.1)$	&	$1.9 (0.8)$	&	$-0.2	$\\
Fe II	&	$5018.434$	&	$1.6  ( 0.2)$	&	$14.9  ( 0.5)$	&	$5.4  (	1.4) $&	$-0.4$	\\
Fe II	&	$5169.030$	&	$0.9  ( 0.1)$	&	$13.4  ( 0.2)$	&	$6.5  (	1.4)	$&	$-0.5$\\
Fe II	&	$5197.577$	&	... & ... & $<1.2$ & $> -0.1$	\\
Fe II	&	$6432.680$	&	... & ... & $<0.1$ & $> -0.01$	\\
\hline

\label{table:UVES_line_prop}
\end{longtable}
\begin{list}{}{}
\item[{Note:}]{The error in the center was calculated considering the contribution of the error in the Gaussian fit and in the stellar radial velocity.}
\end{list}
}

\newpage
\Online

\begin{appendix}

\section{Additional Material}

\subsection{VISIR transmission spectra}
\label{section:transmission_spec}

\begin{figure*}[!ht]
\begin{center}
\includegraphics[width=1.\textwidth]{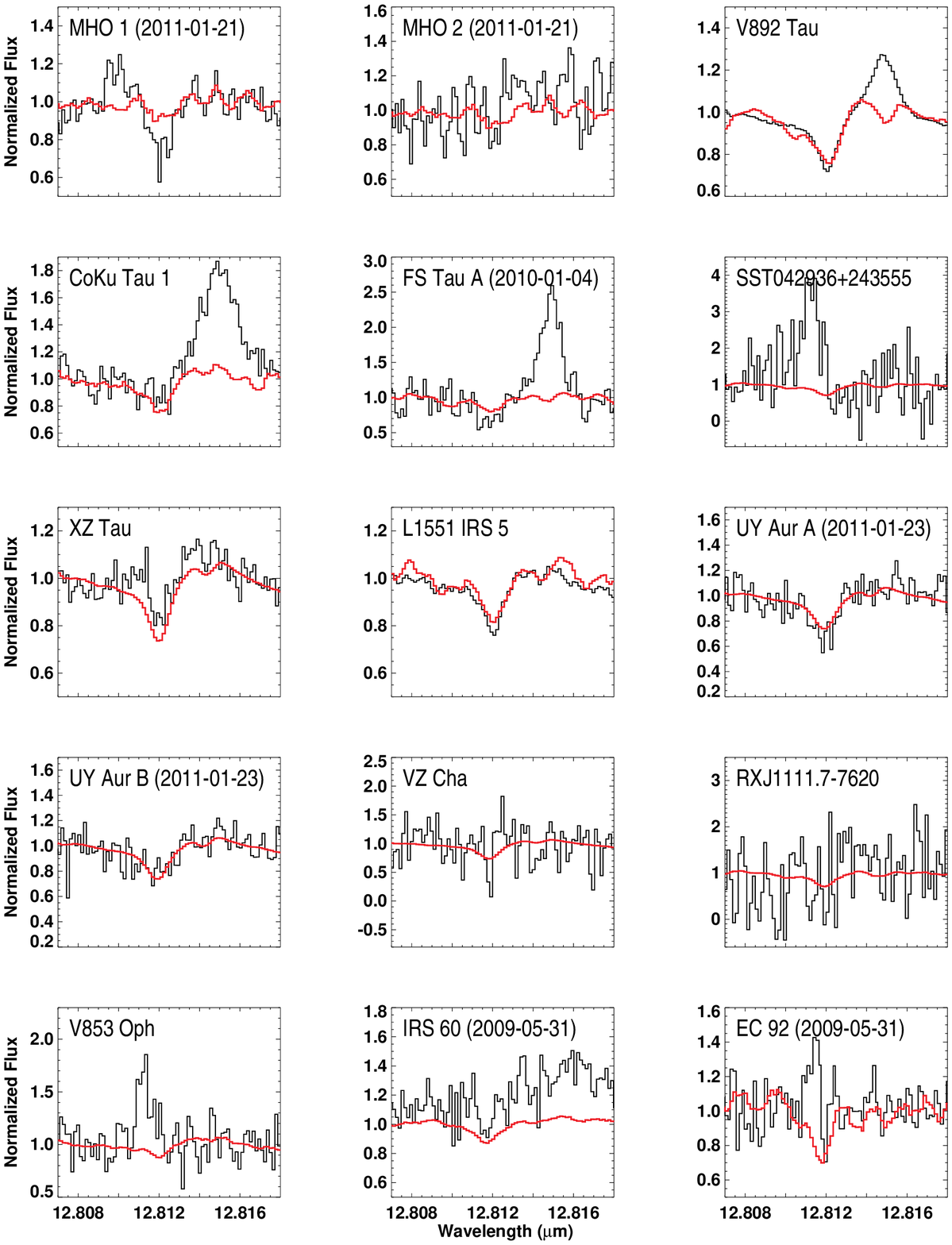}
\end{center}
\caption{VISIR transmission spectra. In black the spectra of the star and in red the spectrum of the telluric standard. We included one spectrum for the stars observed in two days. The spectra are presented before the barycentric and radial velocity correction to visualize the telluric absorption lines and the observed science and standard star/asteroid spectra.}
\label{fig:VISIR_trans}
\end{figure*}

Figure~\ref{fig:VISIR_trans} provides the VISIR transmission spectra for both our targets and the telluric standard.

\subsection{Radial velocities used in this study}
\label{section:radial_vel}
\begin{table*}[!ht]
\begin{center}
\caption{Radial velocities obtained from UVES spectra}
\begin{tabular}{lcccc}
\hline
		Name & $v~ \sin i $& v$_{\rm radial}$  &   v$_{\rm radial}$ & \\
               &  (km~s$^{-1}$) &   (km~s$^{-1}$) &(km~s$^{-1}$)  & \\
& This work & This work & Literature & Reference  \\
\hline
CoKu~Tau~1  & $ 10~(3)$ & ${17.5}~ (2.4)$ & ${15.0 (0.8)}$ & {\citet{white:2004aa}} \\ 
FS~Tau~A    & $15~(5)$ &$17.1~ (0.1)$ & $24~ (7) $ & \citet{folha:2000aa}\\  
		    &                & & $18~ (5)$ & \citet{eisloffel:1998aa}\\
V853 Oph		& $<~7$ & $-8.9~ (0.1)$ & ... & \\ 
\hline
\end{tabular}
\label{uves_radial}
\end{center}
\end{table*}

 Since the stellar radial velocity is a key parameter in determining the centroid of the emission lines, it is worth to explain the choice of radial velocities used in this study.

For the stars located in the Taurus region with no measurement of the radial velocity available in the literature, we adopted the average radial velocity of Taurus, $v_{\rm rad} =16.0 \pm 0.4$~\kms~\citep{bertout:2006aa}. We also use this value for the Herbig Be star V892 Tau. The $^{12}$CO $J=2-1$ observations by \citet{panic:2009ab} were likely dominated by extended emission from the cloud. A search through the \citet{herbig88} catalogue within $1'$ of V892 Tau for radial velocities of good quality also argued in favor of a radial velocity around $15-16$~\kms~ for V892 Tau.

In addition, the radial velocities used for two stars (L1551~IRS5 and EC~92) are reported in the literature using the frame of the Local Standard of Rest (LSR), $v_{\rm LSR}$ \citep{covey:2006aa}. In these cases the radial velocities were converted to $v_{\rm rad}$ using the expression ${\rm v_{rad}}={\rm v_{LSR}}-\left(U_\odot \cos l + V_\odot \sin l \right) \cos b -W_\odot$, where  $l$ and $b$ are the coordinates of the star in the Galactic coordinate system, and $U_\odot=10.3$~\kms, $V_\odot=15.3$~\kms, and $W_\odot=7.7$~\kms~  come from \citet{niinuma:2011aa}.

\subsubsection{Determination of the stellar radial velocity from UVES spectra}
\label{app:rad_vel_cal}

To determine the radial velocity of our program stars, we used the optical spectra obtained by UVES. 
Optical spectra are populated by a series of absorption lines from the stellar photosphere, where the lines are shifted due to the stellar motion with respect to the observer, the radial velocity.
In order to determine the line shift, each spectrum was compared with a high-resolution synthetic spectrum \citep{bertone:2008aa} of a star having the same spectral type as the star observed.
Where an appropriate synthetic spectrum was not available, we used a synthetic spectrum of similar spectral type as the star observed. 
For this purpose we used the interactive IDL-based software described in \citet{carmona:2010aa}.
This tool allows the synthetic and target spectra to be displayed, and to calculate the radial velocity of the observed spectrum using the cross-correlation technique.
The maximization of the cross-correlation function is a widely-used procedure to determine radial velocities (e.g.,~\citealt{tonry:1979aa,allende-prieto:2007aa}).
If we have two arrays, $S$ corresponding to the stellar spectrum and $T$ the template spectrum, the cross-correlation between the two will be a new array $C$ defined by the following expression:

\begin{equation}
C_i = \sum_k T_k~S_{k+i}
\end{equation}

The maximum value of the cross-correlation function $C$ will correspond to the element $i=p$ where $p$ is the shift in pixels between both the stellar spectra $S$ and the template spectrum $T$.  
The position of the maximum of the function (the radial velocity) and its error were calculated with a Gaussian fit. 
The radial velocity measured by cross-correlation was corrected by the barycentric motion of the earth at the moment of the observations to obtain the final radial velocity measurement.
The radial velocities obtained with this method are presented in Table~\ref{uves_radial}, where we also included the radial velocities from the literature, when available. Errors are reported between parentheses.
For one object (V853~Oph) there is no previous measurement of the radial velocity, while for FS~Tau~A there are two measurements available in the literature, both having uncertainties larger than $5$~km~s$^{-1}$. 
For FS~Tau~A and V853~Oph the precision achieved in this work is $<~1$~\kms.
Finally, due to the lower signal-to-noise ratio obtained in the spectrum of CoKu Tau~1, fewer photospheric lines were detected. Therefore, the precision achieved in the radial velocity is lower for this star ($17.5 \pm 2.4$~\kms) {than that derived by \citet{white:2004aa}, although our value is consistent. Therefore we used their radial velocity in this paper.}

 In addition, we obtained the $v\sin~i$ for the three stars by comparing the observed spectra with rotationally broadened synthetic spectra (e.g., \citealt{bertone:2008aa,muller:2011aa}) of $T_{\rm eff}= 3850$~K and $\log g = 4.5$ for CoKu Tau 1 and FS Tau A. For V853~Oph the values used were $T_{\rm eff}= 3370$~K  and $\log g = 4.5$, the lines are spectrally unresolved. The results are reported in Table~\ref{uves_radial}.

\subsection{Lines detected in UVES spectra}
\begin{figure*}
\begin{center}
\includegraphics[width=\textwidth]{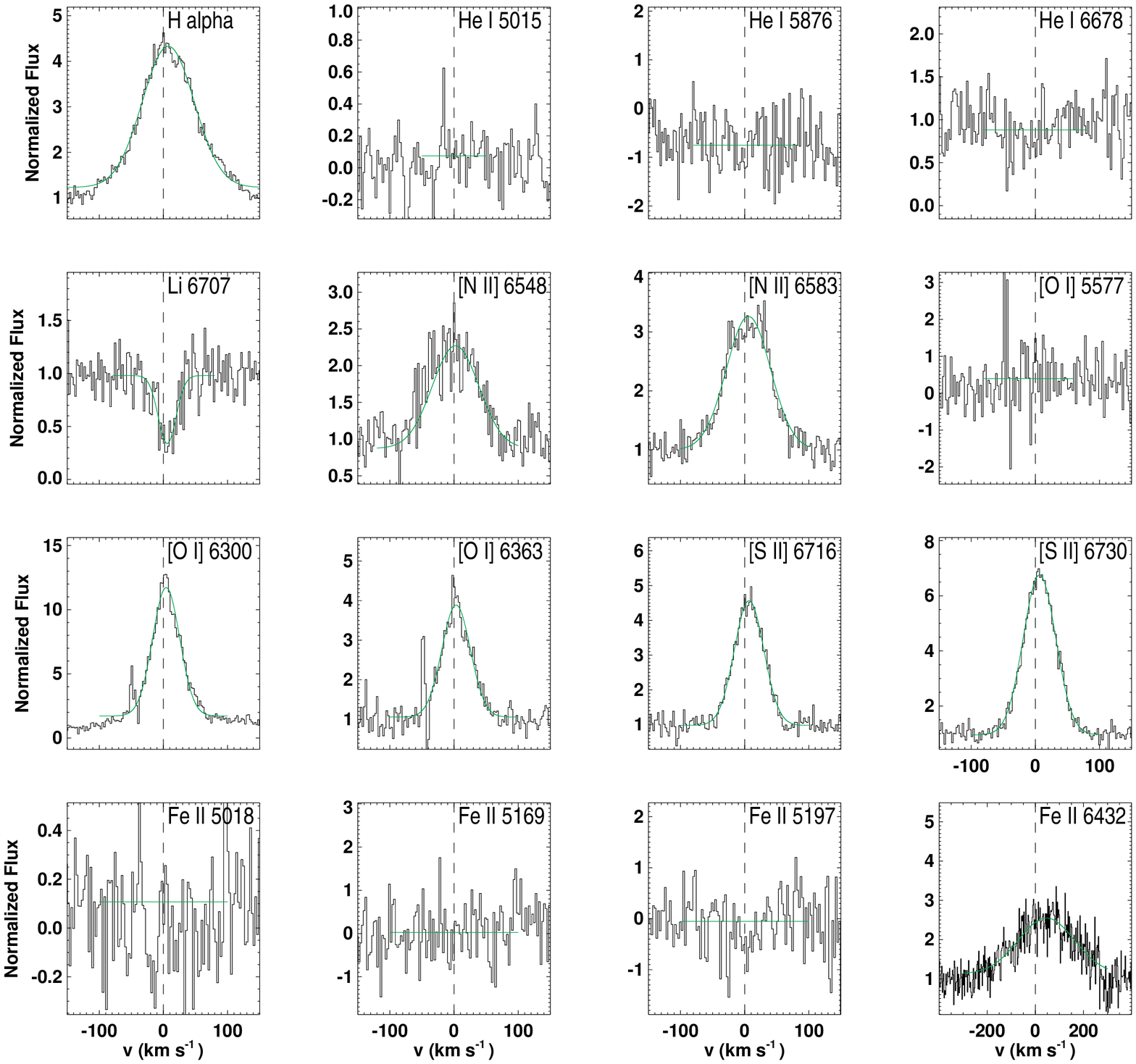}
\end{center}
\caption[Lines detected in the UVES spectrum of CoKu Tau 1]{Lines detected in the UVES spectrum of CoKu Tau 1. The spectrum was normalized to the continuum level. The fitted Gaussian profile is overplotted in green solid line. }
\label{fig:CoKU_optical_lines}
\end{figure*}

\begin{figure*}
\begin{center}
\includegraphics[width=\textwidth]{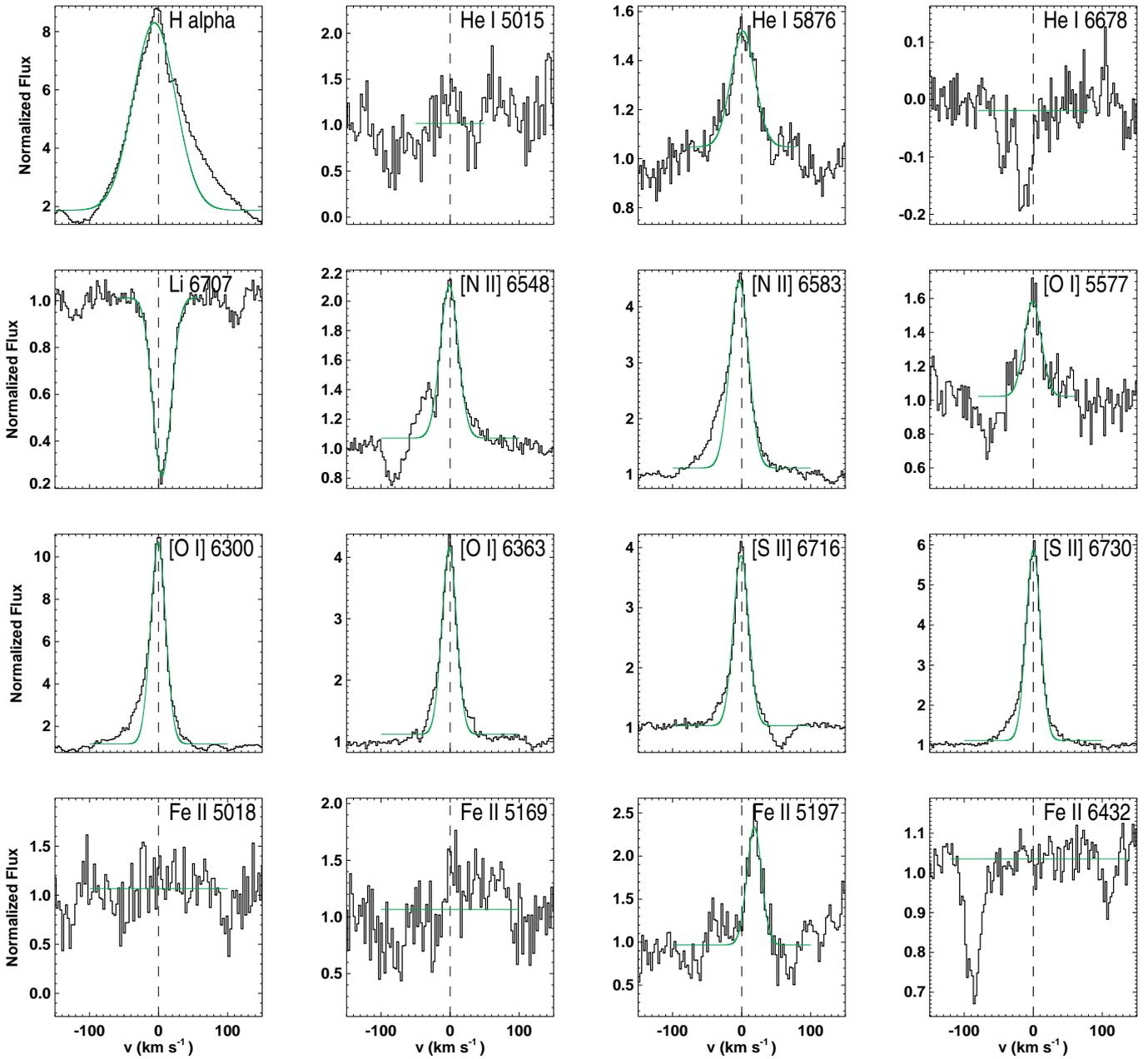}
\end{center}
\caption[Lines detected in the UVES spectrum of FS Tau A]{Lines detected in the UVES spectrum of FS Tau A. The absorption feature next to [S~II] ($\lambda 6716$) is the absorption line of Ca~I ($\lambda 6717$).}
\label{fig:FSTAU_optical_lines}
\end{figure*}

\begin{figure*}
\begin{center}
\includegraphics[width=\textwidth]{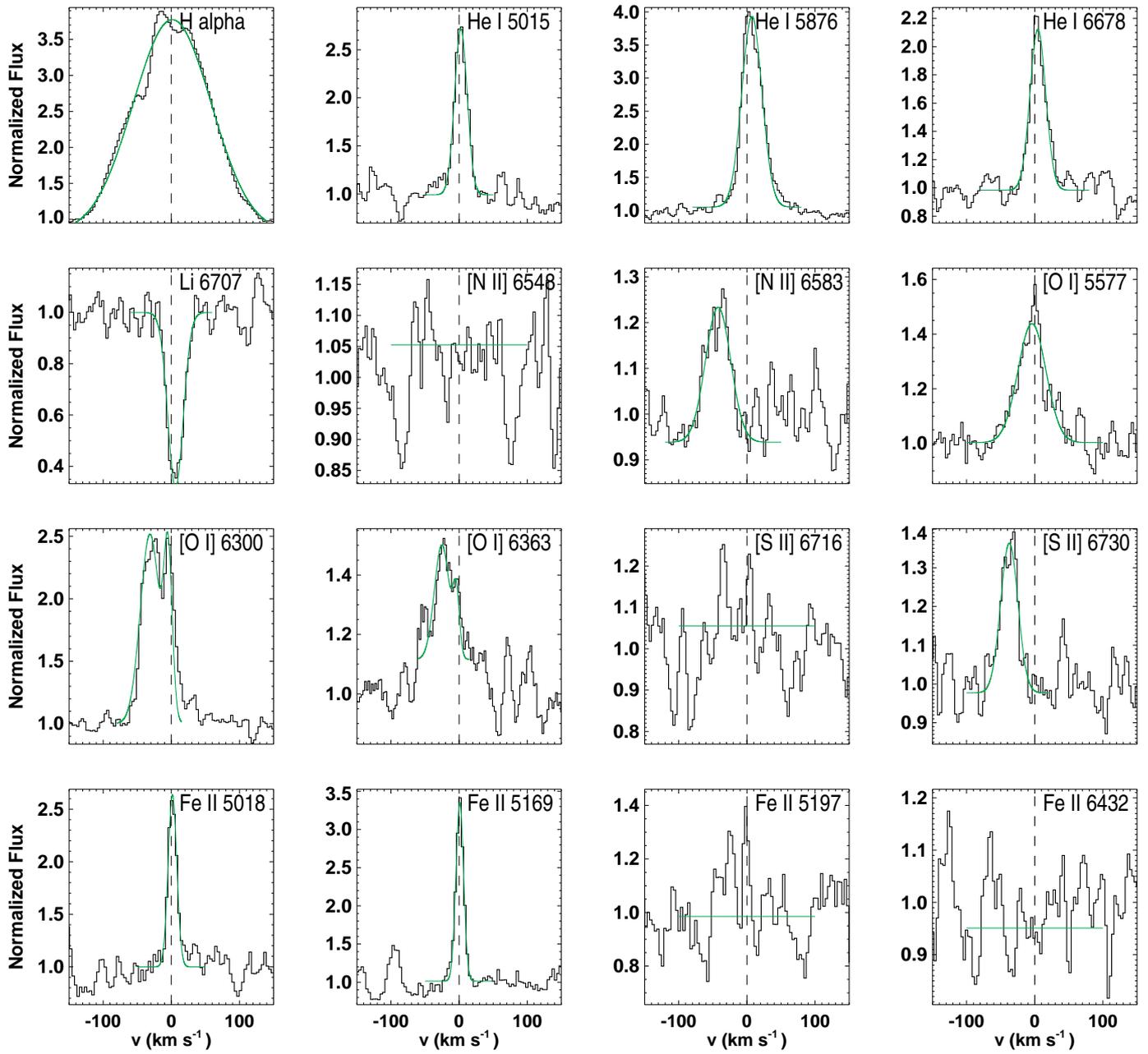}
\end{center}
\caption[Lines detected in the UVES spectrum of V853 Oph]{Lines detected in the UVES spectrum of V853 Oph. The profile of \oxy~$\lambda 6300$ was fitted with two Gaussians to account for the high and low-velocity components observed.}
\label{fig:V853OPH_optical_lines}
\end{figure*}

\end{appendix}

\end{document}